\documentclass[10pt, reqno]{amsart}
\usepackage{amsmath, amsthm, amscd, amsfonts, amssymb, graphicx, subcaption, color, multirow}
\usepackage[bookmarksnumbered, colorlinks, plainpages]{hyperref}

\makeatletter \oddsidemargin.9375in \evensidemargin \oddsidemargin
\def\authorsaddresses#1{\dedicatory{#1}}
\newtheorem{theorem}{Theorem}[section]

\theoremstyle{definition}
\newtheorem{definition}[theorem]{Definition}

\theoremstyle{remark}

\numberwithin{equation}{section}

\begin{document}


\title[Dependence control chart using maximum copula entropy]{Dependence control chart using maximum copula entropy}

\author[Fallah et al.]{Seyedeh Azadeh Fallah Mortezanejad$^{1}$, Ruochen Wang$^{*2}$, Gholamreza Mohtashami Borzadaran$^3$, Kim Phuc Tran$^{4,5}$}

\authorsaddresses{$^{1,2}$School of Automotive and Traffic Engineering, Jiangsu University, Jiangsu, China.\\
$^{3}$ Department of Statistics, Faculty of Mathematical Sciences Ferdowsi University of Mashhad, Mashhad, Iran.\\
$^{4}$Univ. Lille, ENSAIT, ULR 2461 - GEMTEX - Génie et Matériaux Textiles, F-59000 Lille, France\\
$^{5}$International Chair in DS \& XAI, International Research Institute for Artificial Intelligence and Data Science, Dong A University, Danang, Vietnam.} 
\subjclass[2010]{Primary 47A55; Secondary 39B52, 34K20, 39B82.}

\keywords{Control chart; Maximum entropy; Copula function; Spearman's rho; $T^2$-Hotelling statistic.}

\begin{abstract}
Statistical quality control methods are noteworthy to producing standard production in manufacturing processes. In this regard, there are many classical manners to control the process. Many of them have a global assumption around the distributions of the process data. They are supposed to be Normal, but it is clear that it is not always valid for all processes. Such control charts made some wrong decisions that waste funds. So, the main question while working with multivariate data set is how to find the multivariate distribution of the data set, which saves the original dependency between variables. To our knowledge, a copula function guarantees dependence on the result function. It is not enough when there is no other fundamental information about the statistical society, and we have just a data set. Therefore, we apply the maximum entropy concept to deal with this situation. In this paper, first of all, we get the joint distribution of a data set from a manufacturing process that needs to be in-control while running the production process. Then, we get an elliptical control limit via the maximum copula entropy. Finally, we represent a practical example using the method. Average run lengths are calculated for some means and shifts to show the ability of the maximum copula entropy. In the end, two practical data examples are presented, and the results of our method are compared with the traditional way based on Fisher distribution.
\end{abstract}

\maketitle


\section{Introduction}
Shannon entropy has been introduced first by \cite{Shannon} since $ 1948 $. Afterwards, it is applied in many different fields. The maximum entropy principle was presented by \cite{Jaynes} since $ 1957 $. Jaynes exerted the Lagrange function according to some constraints to find the maximum entropy distribution. Such papers as \cite{Kagan, Shore, Johnson} studied the maximum entropy concept. After that, it is used wieldy by authors until recent years as \cite{Cesari, Fallah, Sutter}. \\
The maximum entropy principle is a good method of finding the unknown distribution of a univariate data set because it does not need any strong presumption about distribution. Working well with ill-posed conditions and not requiring large sample sizes makes it a suitable choice. Although all benefits of the maximum entropy concept, it can be difficult for some researchers to define more constraints for multivariate data sets to preserve the original dependency between different variables of multivariate data. A specialist needs to save it in the result distribution function. Some papers like \cite{chu2011recovering, zhao2011copula, piantadosi2012copulas, Chen2014, Singh, Rahmani, Mortezanejad} linked the maximum entropy principle and copula function. Generally, by the aim of both concepts, we get a copula density function by a maximum copula entropy via adding some simple constraints according to intended dependency measures. So, the maximum copula function has the same dependency on the existing data. Finally, the Sklar theorem \cite{Sklar} helps easily to get the multivariate distribution function whose dependency is the same as the available data. \\
In this paper, we would like to peruse manufacturing process data which generates multivariate data sets with unknown distributions. Typically they are assumed to be Normal distributions. This assumption is incorrect in general cases. So, technical assistants need to know the distribution. In this regard, the main point is to transfer the original dependency to the result density function. Thus, we are working on this issue to combine the maximum entropy principle and copula function. As mentioned, the maximum entropy principle is applied to find the empirical multivariate distribution, and the copula function cares about the dependency. Our predestinate data is bivariate and also dependent. So, we estimated its distribution by the maximum entropy principle for some simulated dependency measures based on Spearman's rho and Blest measures. In the next step, we apply the $T^2$-Hotelling statistic, which is common for dealing with a multivariate data set. Afterwards, we compute the statistical quality control for these kinds of data. These control limits are reliable because the dependency is paid attention to while calculating them. \\
The proceeds of this paper are: In section \ref{CopulaSec}, some basic concepts of copulas and dependent measures are determined. In section \ref{EntropySec}, we present univariate and bivariate distributions with the calculation procedure. The purpose of the bivariate case is to compare functionally with the result function of the next section. In section \ref{sec2}, we explain the procedure of finding a bivariate maximum entropy distribution concerning some intended constraints. Then, we clarify the maximum copula entropy according to corresponding conditions. We apply Shannon entropy for both cases. In the process of acquiring the maximum copula entropy, we exert some dependence measures to transfer the dependency of an available data set to the final function. In the following section, we use the maximum copula entropy to get the joint density function of the data set using the Sklar theorem. In section \ref{sec3}, we represent the $ T^2 $-Hotelling statistic and illustrate how to find the statistical control limits for bivariate data set with its original dependency saved by the maximum copula function and describe how to compute $ARL$s. In section \ref{Prac}, we calculate the coefficients of the maximum copula entropy for some instance values of dependence measures along with their surface plots shown in the figures. Then, we estimate the upper control limit for some different means with their corresponding $ARL$s. In section \ref{RealSec}, two practical data examples are discussed in detail.
In section \ref{sec5}, we make a conclusion and statements of the paper.
\section{Copula function definition}\label{CopulaSec}
Statistics researchers consider copula functions as an advanced method of dealing with data sets. It has many beneficial properties in saving dependency of the data to raise the precision. For example, the copula function is applied to generate random variate from a set with the same dependency whose distribution is unknown. Sklar introduced copula functions since $1959$ in \cite{Sklar} applying one-dimensional marginal functions to build multivariate distributions. Fisher $(1997)$ \cite{Fisher1997} was the first published paper in statistics using the copula function, and also Schweizer and Wolff \cite{Schweizer1981} are the pioneers. After that, this concept was wieldy used in many different papers like \cite{Liu2018, Wang2019, Zhang2019}. In this paper, we focus on the copula function mixed with the entropy principle, \cite{Cerqueti2018, Jian2019, Mortezanejad, Wang2020}. The idea is to link the copula function and the maximum entropy to estimate the unknown copula and then approximate the indistinct distributions using the Sklar theorem \cite{Sklar}. We add dependence measures to save the dependency in the data set, but the copula function requires their values. So, some pre-estimated statistics have to be defined. We present some primary definitions and theorems. The copula description is in the following, \cite{Nelsen2006}:
\begin{definition}
A two-dimensional copula is a function defined on $I^2$ where $I=[0,1]$ with the following properties:
\begin{itemize}
  \item for every $u,v \in I$: $$C(u,0)=C(0,v)=0,~C(u,1)=u,~C(1,v)=v,$$
  \item for all $u_1$, $u_2$, $v_1$, and $v_2$ in $I$ such that $u_1\leq u_2$ and $v_1\leq v_2$: $$C(u_2,v_2)+C(u_1,v_1)-C(u_2,v_1)-C(u_1,v_2)\geq 0.$$
\end{itemize}
\end{definition}
In multivariate data studying, copula functions have a valuable rule based on the Sklar theorem.
\begin{theorem}
Let $H(\cdot,\cdot)$ be a joint distribution function for random variables $X$ and $Y$ whose marginal functions are $F_X(\cdot)$ and $F_Y(\cdot)$. Then a copula $C(\cdot,\cdot)$ exists such that
\begin{equation}\label{copula}
  H(x,y)=C(F_X(x),F_Y(y)), \forall x,y\in \mathbb{R}.
\end{equation}
$C(\cdot,\cdot)$ is unique if $F_X(\cdot)$ and $F_Y(\cdot)$ are continuous; otherwise, $C(\cdot,\cdot)$ can be uniquely defined on the joint support set $\mathbb{S}(X,Y)$. Reciprocally, let $C(\cdot,\cdot)$ be a copula function, and $F_X(\cdot)$ and $F_Y(\cdot)$ be univariate distribution functions. Then $H(\cdot,\cdot)$ in \eqref{copula} is the corresponding joint distribution function respect to the margins.
\end{theorem}
The key in the copula topic is this theorem, which is applied in several articles with different issues. The theorem gives us a connection between the copula function and the joint distribution function. In the following paper, some dependence measures are exerted. The first measure is Spearman's rho evaluating coordination and incoordination between variables. Spearman's rho definition is based on \cite{Kruskal1958}. Let $ (X_1,Y_1) $, $ (X_2,Y_2) $, and $ (X_3,Y_3) $ be three vectors of independent random variables whose joint distribution function is $H(\cdot,\cdot)$ with margins $F_X(\cdot)$ and $F_Y(\cdot)$ and their corresponding copula function is $C(\cdot,\cdot)$. The Spearman's rho is determined by the below formula whose domain is in $[-1,1]$:
\begin{equation*}
\rho = 3 \{ P((X_1-X_2)(Y_1-Y_3)>0)-P((X_1-X_2)(Y_1-Y_3)<0) \}.
\end{equation*}
The following theorem in \cite{Nelsen2006} applied copula function to compute Spearman's rho.
\begin{theorem}
Suppose $ X $ and $ Y $ be two independent random variables with copula $C(\cdot,\cdot)$. Then, Spearman's measure is calculated by:
\begin{eqnarray*}
\rho &=& 3Q(C,\Pi) \\
&=& 12 \int_{I^2} uv dC(u,v) -3	 \\
&=& \int_0 ^1 \int_0 ^1 C(u,v) du dv -3.
\end{eqnarray*}
\end{theorem}
The other used measures are related to Blest rank correlations adapted from \cite{Blest2000} called the first, second, and third Blest's measures:
\begin{eqnarray*}
\nu_1 &=& 2-12 \int \int_{I^2} (1-u)^2 v c(u,v) du dv, ~~\nu_1\in [-1,1],\\
\nu_2 &=& 2-12 \int \int_{I^2} u(1-v)^2 c(u,v) du dv, ~~\nu_1\in [-1,1],\\
\eta &=& 6 \int \int_{I^2} u^2 v^2 c(u,v) du dv-\frac{1}{5}, ~~\eta \in [0,1].
\end{eqnarray*}
As we can see, the density copula function is required for all presented scales. If we work with a real data set, then the distribution and copula function are unknown. A logical recommendation is to put pre-estimators based on the sample data. The primary aim of this paper is to appraise the copula and joint distribution function. Paper \cite{chu2011recovering} determined the estimations in the article as well, but before the presentation, it needs to define some notations. $n$ is the sample size, and $u_n$ and $v_n$ are defined for $t=1,\ldots,n$:
\begin{eqnarray*}
  u_t &=& \frac{1}{n}\sum_{i=1} ^n 1(X_i\leq X_t)\hspace{0.23cm}=\hspace{0.23cm}\frac{R_t}{n+1}, \\
  v_t &=& \frac{1}{n}\sum_{i=1} ^n 1(Y_i\leq Y_t)\hspace{0.36cm}=\hspace{0.36cm}\frac{S_t}{n+1},
\end{eqnarray*}
where $1(\cdot)$ is the indicator function. The pre-estimation of explained dependencies are:
\begin{eqnarray*}
  \widehat{\rho} &=& \frac{12}{n^3-n}\sum_{t=1} ^n R_tS_t-3\frac{n+1}{n-1}, \\
  \widehat{\nu_1} &=& \frac{2n+1}{n-1}-\frac{12}{n^2-n}\sum_{t=1} ^n \left(1-\frac{R_t}{n+1}\right)^2S_t, \\
  \widehat{\nu_2} &=& \frac{2n+1}{n-1}-\frac{12}{n^2-n}\sum_{t=1} ^n R_t \left(1-\frac{S_t}{n+1}\right)^2, \\
  \widehat{\eta} &=& \frac{6}{n^2-n} \sum_{t=1} ^n \left(\frac{R_t}{n+1}\right)^2\left(\frac{S_t}{n+1}\right)^2-\frac{(1/5)n+1}{n-1}.
\end{eqnarray*}
By the copula and entropy principle, we can find out the unclear distribution of any set in the following.
\section{Maximum entropy principle}\label{EntropySec}
The entropy was introduced by Shannon in \cite{Shannon, Shannon1949}. The Shannon entropy is applicable in statistics and broadly used in many other fields like mathematics, physics, computer science, economics, etc. Jaynes explained the maximum entropy principle in \cite{Jaynes} since $1957$, which has many advantages like being unbias, suitable for small sample sizes, no need for strong assumptions, etc. The maximum entropy concept is an applied way to find the unknown distribution of a data set. It gets a compatible distribution for available information. In this manner, we use the maximum entropy principle to approximate margins and joint distribution. \\
In the following, we describe how to find univariate maximum copula entropy according to the Shannon, which is useful for getting marginal functions. Then, the bivariate function is presented. The Shannon entropy for random variable $X$ in the continuous case is the differential entropy as follows:
\begin{equation*}
  \mathcal H_S (f_X) = \int_{\mathbb{S}_X} -\log f_X(x) dF_X(x),
\end{equation*}
where $\mathbb{S}_X$ is univariate support set, and $f_X(x)$ is its density function. The next step is to add some constraints. Kagan et al. \cite{Kagan} extended some conditions on the entropy. So, some mandatory and optional constraints have to be defined on the univariate density function:
\begin{displaymath}
\left\{\begin{array}{ll}
\int_{\mathbb{S}_X} d F_X(x)=1, \\
E(g_i(X))=m_i(x),~ j=1,\ldots,k, \\
\end{array}\right.
\end{displaymath}
where $k$ is the number of optional constraints, $m_i(x)$ for $j=1,\ldots,k$ are known based on the available data whose corresponding functions are $g_i(X)$. The first condition guarantees the result is a valid statistical density. The Lagrange function for this case is:
\begin{eqnarray}\label{L}
L(f_X,\lambda_0,\ldots,\lambda_k)=&-&\int_{\mathbb{S}_X} \log f_X(x) dF_X(x)-\lambda_0 \{ \int_{\mathbb{S}_X} dF_X(x)-1\}   \\
&-&\Sigma_{i=1} ^k \lambda_i \{ \int_{\mathbb{S}_X} g_i(x,y) dF_X(x)-m_i(x)\}. \nonumber
 \end{eqnarray}
After differentiating and setting it to zero, the final univariate maximum entropy is as below:
\begin{equation*}
  f_X(x) = \exp (-\lambda_0-\Sigma_{i=1} ^k \lambda_i g_i(x)),~~ x\in \mathbb{S}_X.
\end{equation*}
So, we briefly expiated how to get the maximum entropy function for the case with one variable. That is helpful in the proceeding of joint distribution function based on copula. It is worth to describe for the bivariate state because we would like to compare functionally the result of pure entropy function with the outcome of the manner combining with copulas. By the way, we will make a joint density function based on the maximum entropy. So, the bivariate form of Shannon entropy is:
\begin{equation*}
  \mathcal H_S (f_{X,Y}) = \int \int _{\mathbb{S}(X,Y)} -\log f_{X,Y}(x,y) dF_{X,Y}(x,y),
\end{equation*}
which is for $ X $ and $ Y $ whose density and distribution function are $ f_{X,Y}(x,y) $, and  $F_{X,Y}(x,y) $ respectively, and $\mathbb{S}(X,Y)$ is the joint support set. Some intended constraints are needed to find the joint maximum entropy distribution as well:
\begin{displaymath}
\left\{\begin{array}{ll}
\int \int_{\mathbb{S}(X,Y)} d F_{X,Y}(x,y)&=1, \\
E(g_i(X,Y))&=m_i(x,y),~ j=1,\ldots,k', \\
\end{array}\right.
\end{displaymath}
where $ m_i(x,y) $s for $ j=1,\ldots,k' $ are some known moments, which are calculated based on the available data set. $ g_i(X,Y) $s for $ j=1,\ldots,k' $ are corresponding functions to $ m_i(\cdot,\cdot) $s. $ k' $ is the number of constraints on moments, which does not have to be equal to $k$. $ dF_{X,Y}(x,y) $ is the full differential of $ F_{X,Y}(x,y) $. Then the maximum entropy distribution is gotten by applying the Lagrange function made of Shannon entropy and its corresponding constraints as well:
\begin{eqnarray}\label{L}
L(f_{X,Y},\lambda_0,\ldots,\lambda_k')=&-&\int_{\mathbb{S}(X,Y)} \log f_{X,Y}(x,y) dF_{X,Y}(x,y)   \nonumber\\
&-&\lambda_0 \{ \int_{\mathbb{S}(X,Y)} dF_{X,Y}(x,y)-1\} \nonumber\\
&-&\Sigma_{i=1} ^{k'} \lambda_i \{ \int_{\mathbb{S}(X,Y)} g_i(x,y) dF_{X,Y}(x,y)-m_i(x,y)\}. \nonumber
 \end{eqnarray}
Then, the Lagrange function should be differentiated for $ f_{X, Y}(\cdot) $, and by using the Kuhn-Tucker method, joint maximum entropy distribution is found:
\begin{equation}\label{JointME}
  f_{X,Y}(x,y) = \exp (-\lambda_0-\Sigma_{i=1} ^{k'} \lambda_i g_i(x,y)),~~ (x,y)\in \mathbb{S}(X,Y).
\end{equation}
In the next section, the copula concept is added to the maximum entropy procedure to make the effect of available dependency on data. Function $f_{X, Y}(x,y)$ is not as reliable as the result of the maximum copula entropy.
\section{Joint distribution function via maximum copula entropy method}\label{sec2}
In this section, we would like to present a feasible method of finding multivariate distribution affected by the dependency between variables. For simplicity of calculations and notations, we discuss the bivariate data set. Although there is the main question while working with a multivariate data set whose distribution is unknown, the maximum entropy seems fine for this purpose. The question is, how to found the distribution with the same original dependency between corresponding variables? The copula function replies to the question as well. Generally, function \eqref{JointME} is upgraded by copula function to preserve the dependency. Here, we suppose to mix these two major concepts to estimate a fit distribution. We are keen on representing how to find the maximum copula entropy. First of all, the copula entropy based on the Shannon definition is:
\begin{equation*}
  \mathcal H_S (c) = \int \int _{I^2} -c(u,v) \log c(u,v) du dv,
\end{equation*}
where
\begin{equation*}
  c(u,v)=\frac{\partial^2C(u,v)}{\partial u \partial v}.
\end{equation*}
The maximum copula entropy has to be found out based on some constraints ensuring the result function is the copula. These essential constraints according to \cite{chu2011recovering} are for $i=1,\ldots,r$:
\begin{displaymath}
\left\{\begin{array}{ll}
\int \int_{I^2} c(u,v)dudv&=1, \\
\int \int_{I^2} u^i c(u,v)dudv&=\frac{1}{i+1}, \\
\int \int_{I^2} v^i c(u,v)dudv&=\frac{1}{i+1}, \\
\end{array}\right.
\end{displaymath}
where $ r $ is the counter of constraints and the bigger choice of $ r $, the more accurate creature of the result function compared with copulas. Here, we would like to add other equations based on some measures of dependence that should be estimated while dealing with real data sets to get a copula function. This copula has the same dependency as the available data set. These constraints are related to $\rho$, $\nu_1$, $\nu_2$, and $\eta$. According to \cite{chu2011recovering, Chu2016}, some phrases have approximately equal Lagrange coefficients. For example, in paper \cite{Mortezanejad}, when they put different coefficients for each constraint in a simulation study, their results were almost the same. It comes from the symmetricity of the maximum copula function. Thus, we incorporate some of them to reduce the Lagrange coefficients. Synchronization is significantly important while dealing with real data because the number of computations reduces saving time and energy. The incorporations are obvious and explained more in the following. After reduction, the optional conditions are:
\begin{equation}
 \begin{cases}
  \int\int_{I^2} uv~ c(u,v)dudv&=\frac{\rho+3}{12}, \\
  \int\int_{I^2} u^2v~ c(u,v)dudv&=\frac{2\rho-\nu_1+2}{12}, \\
  \int\int_{I^2} u^2v^2~ c(u,v)dudv&=\frac{\eta + \frac{1}{5}}{6},
  \end{cases} \label{measured}
\end{equation}
where $ \rho $, $ \nu_1 $, and $ \eta $ are Spearman's rho, Blest measures $ I $, and $ III $, respectively. It is worth mentioning that the value of \eqref{measured} is Blest measure $ II $. To find the maximum copula entropy, we have to apply the Lagrange function and Kuhn-Tucker method as well as before, and the result copula function is:
\begin{align}\label{MaxCopulaE}
c(u,v) = \exp &\left( -1-\lambda_0-\Sigma_{i=1} ^r \lambda_i(u^i+v^i) \right. \\ \nonumber
& \left. -\lambda_{r+1} uv -\lambda_{r+2} (u^2v+uv^2)-\lambda_{r+4}u^2v^2\right),~\forall u,v\in [0,1].
\end{align}
The values of $ \lambda_i $ for $i=0,\ldots,r+4$ are gotten by applying $ c(u,v) $ in the intended constraints, and a system of equations has to be solved. In practice, dependence measures must be estimated first, but the copula function is required in their computations. In section \ref{CopulaSec}, some pre-estimators are determined to solve this problem. We find copula functions under some estimated dependence measures in section \ref{Prac}. By the way, after getting the copula density function related to the dependence measures, their joint density functions can be obtained by this formula:
\begin{equation}\label{fco}
  f_{X,Y}(x,y)=c(u,v)f_X(x)f_Y(y),
\end{equation}
where $ f_X(\cdot) $ and $ f_Y(\cdot) $ are marginal functions gotten by the maximum entropy principle based on Shannon's definition. In this regard, the functions \eqref{JointME} and \eqref{fco} can be functionally compared. Although \eqref{JointME} has no effect of dependency, the \eqref{fco} highly performs the dependency in data, which was our aim. \\
So, the joint density function of dependence data is gotten via the maximum entropy and copula function. The maximum entropy principle is applied because it is the best choice for the lack of details. It helps us to fit a distribution when there is incomplete information, and the sample size is not large enough. The copula function keeps the original dependency between variables of the data set. Thus, the result of the joint density function is reasonable for our goal. In the next section, we present $T^2$-Hotelling statistics because control limits are designed for multivariate cases.
\section{New control chart using $T^2$-Hotelling}\label{sec3}
In the previous section, we had some estimated dependence measures. We got the unknown joint density function of a data set. The goal is to work on data sets obtained from a manufacturing process, and we would like to control the process over time. We need the appropriate statistical control limits $ [LCL, UCL] $ to control the production process. These limits depend on the joint density function of the process and are generally unknown in practice. Some classical methods exist that have a strong assumption on their distribution. The distribution is supposed to be Normal, which is overall for every data received from any production process. This assumption is invalid in many procedures. So, their control limits are affected by a false distribution. So, a decision based on such limits is wrong and wastes many funds. The purpose of this paper is to use the density function \eqref{fco} to compute the proper control limits. It has some advantages mentioned in the following. First, it is not a general distribution for all processes and is gotten separately for each. Second, the data dependency is considered in it, which is its superiority concerning \eqref{JointME}. For this aim, we suppose to apply $ T^2 $-Hotelling statistic to deal with a multivariate data set. We need the joint density function represented in \eqref{fco} to find the proper control limits. First of all, we present the $ T^2 $-Hotelling statistic for a random vector $ \underset{\sim}{X} $ with mean vector $ \underset{\sim}{\mu} $ and variance-covariance matrix $ \Sigma $ as:
\begin{equation*}
  T^2_{Hotelling}=(\underset{\sim}{X}-\underset{\sim}{\mu})' \Sigma ^{-1} (\underset{\sim}{X}-\underset{\sim}{\mu}).
\end{equation*}
In our case of study, we have:
\begin{displaymath}
\left\{\begin{array}{ll}
\underset{\sim}{X} &=\begin{pmatrix}
           X \\
           Y
    \end{pmatrix}, \\
\underset{\sim}{\mu} &=\begin{pmatrix}
           \mu_X \\
           \mu_Y
    \end{pmatrix}, \\
    \Sigma ^{-1} &= \begin{bmatrix}
           a_{11} & a_{12} \\
           a_{21} &  a_{22}
\end{bmatrix}
\end{array}\right.
\end{displaymath}
where $ a_{12} = a_{21} $. It is obvious that $ T^2 $-Hotelling is a positive statistic, and the corresponding $ LCL $ is $ 0 $ because it measures the distance. So, the lower the value, the closer the quality is to the standards. Therefore, we have to solve this equation to get the $ UCL $:
\begin{equation}\label{InequUCL}
    P( T^2_{Hotelling} \leqslant UCL) \geq 1-\alpha,
\end{equation}
where $ \alpha $ is the first type of error and negligible. Then, we have:
\begin{eqnarray*}
  1-\alpha &\leq&   P((\underset{\sim}{X}-\underset{\sim}{\mu})' \Sigma ^{-1} (\underset{\sim}{X}-\underset{\sim}{\mu})\leqslant UCL)\\
    &=& P(a_{11}(X-\mu_X)^2+a_{22}(Y-\mu_Y)^2+2a_{12}(X-\mu_X)(Y-\mu_Y)\leqslant UCL)\\
    &=& {\int \int}_{ \{(x,y)| a_{11}(X-\mu_X)^2+a_{22}(Y-\mu_Y)^2+2a_{12}(X-\mu_X)(Y-\mu_Y) \leqslant UCL\} } f_{X,Y}(x,y) dx~dy.
\end{eqnarray*}
We need the value of $ UCL $ to satisfy the last equation. These control limits are based on the dependency of two variables $ X $ and $ Y $, whose dependence reflects on $ f_{X,Y}(\cdot,\cdot) $. In statistical quality control is common to use average run lengths $(ARL)$to show the performance of the limits. There are two types: first, $ARL_0$ based on the first type of error $\alpha$ meant the number of samples to be taken from the process to see an out-of-control sample unit under a controlled situation:
\begin{equation*}
  ARL_0=\frac{1}{\alpha}.
\end{equation*}
The second $ARL_1$ is according to the second type of error $\beta$ defined as the number of samples taken under controlled conditions until one selection is outside of the control range:
\begin{equation*}
  ARL_1=\frac{1}{1-\beta}.
\end{equation*}
Note that their distributions are geometric. In the next section, we use this method to find the statistical control limits and their $ARL$s for a simulation study.
\section{Simulation example of a manufacturing process}\label{Prac}
\begin{figure}[t!]
    \centering
    \begin{subfigure}[t]{0.5\textwidth}
        \centering
        \includegraphics[height=2in]{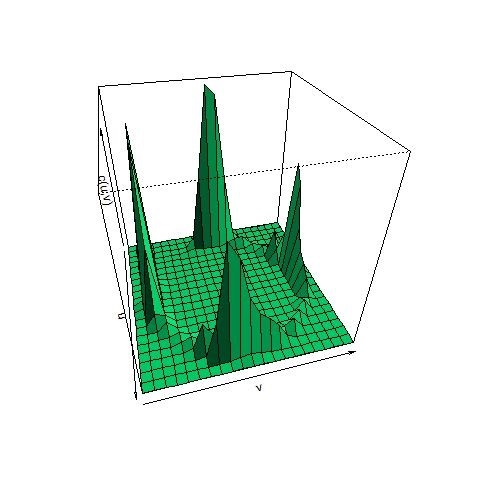}
        \caption{$\rho=-0.4$, $\nu_1=-0.5$, and $ \eta=0.2$}
    \end{subfigure}%
    ~
    \centering
    \begin{subfigure}[t]{0.5\textwidth}
        \centering
        \includegraphics[height=2in]{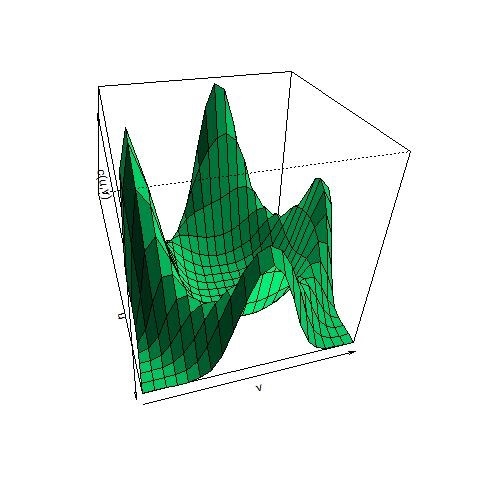}
        \caption{$\rho=-0.1$, $\nu_1=-0.18$, and $ \eta=0.45$}
    \end{subfigure}%
    ~
    \newline
     \centering
    \begin{subfigure}[t]{0.5\textwidth}
        \centering
        \includegraphics[height=2in]{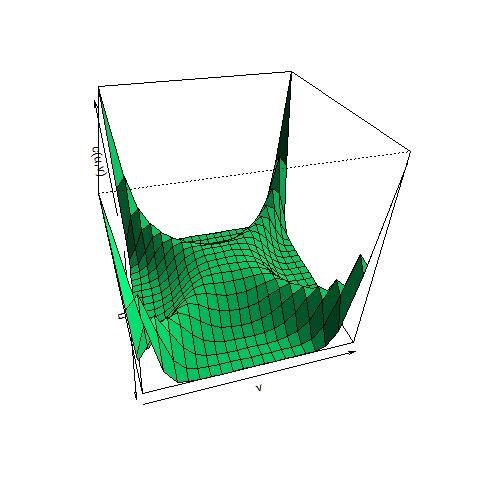}
        \caption{$\rho=0$, $\nu_1=0$, and $ \eta=0.5$}
    \end{subfigure}%
    ~
     \centering
    \begin{subfigure}[t]{0.5\textwidth}
        \centering
        \includegraphics[height=1.75in]{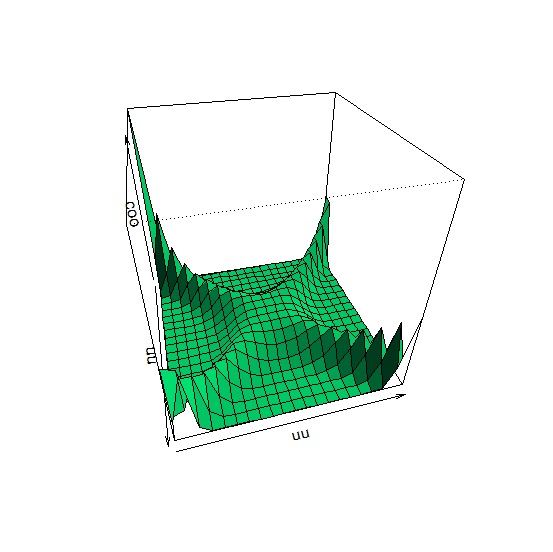}
        \caption{$\rho=0.1$, $\nu_1=0.18$, and $ \eta=0.55$}
    \end{subfigure}%
    ~
    \newline
    \begin{subfigure}[t]{0.5\textwidth}
        \centering
        \includegraphics[height=2in]{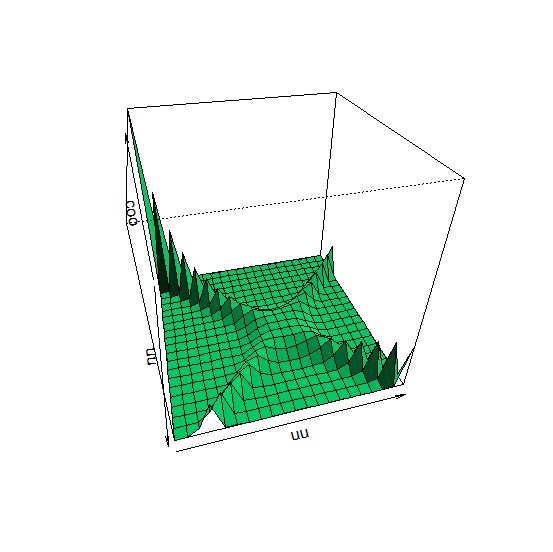}
        \caption{$\rho=0.4$, $\nu_1=0.5$, and $ \eta=0.8$}
    \end{subfigure}
    \caption{Surface plots of Copula density functions for Table \ref{coefco}}\label{plotmed}
\end{figure}

\begin{figure}[t!]
    \centering
    \begin{subfigure}[t]{0.5\textwidth}
        \centering
        \includegraphics[height=2in]{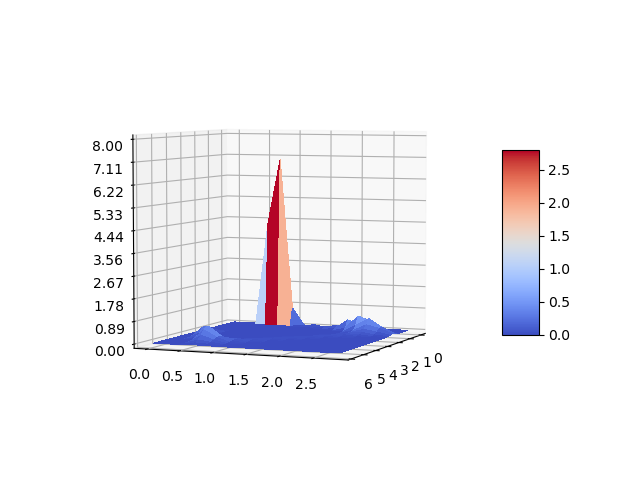}
        \caption{$\mu_X=2$, $\mu_Y=1$ for group $1$}
    \end{subfigure}%
    ~
    \centering
    \begin{subfigure}[t]{0.5\textwidth}
        \centering
        \includegraphics[height=2in]{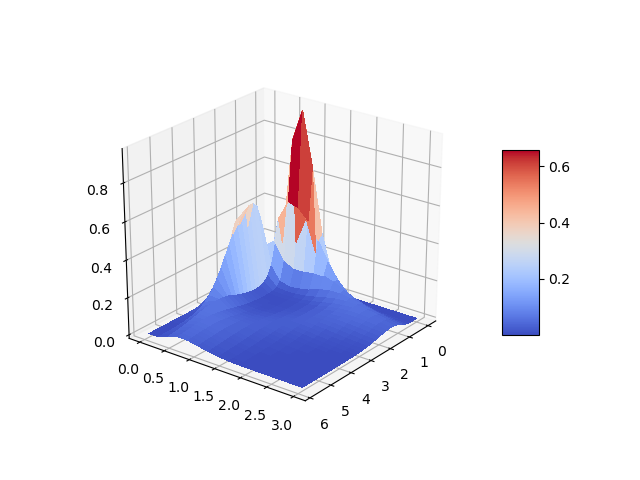}
        \caption{$\mu_X=2$, $\mu_Y=1$ for group $2$}
    \end{subfigure}%
    ~
    \newline
     \centering
    \begin{subfigure}[t]{0.5\textwidth}
        \centering
        \includegraphics[height=2in]{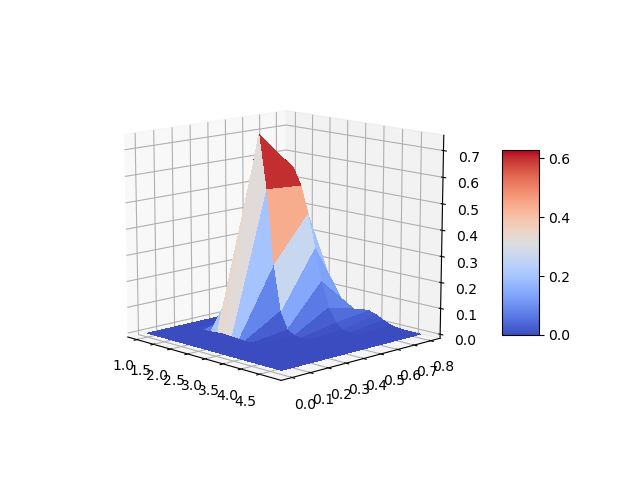}
        \caption{$\mu_X=3$, $\mu_Y=5$ for group $1$}
    \end{subfigure}%
    ~
     \centering
    \begin{subfigure}[t]{0.5\textwidth}
        \centering
        \includegraphics[height=2in]{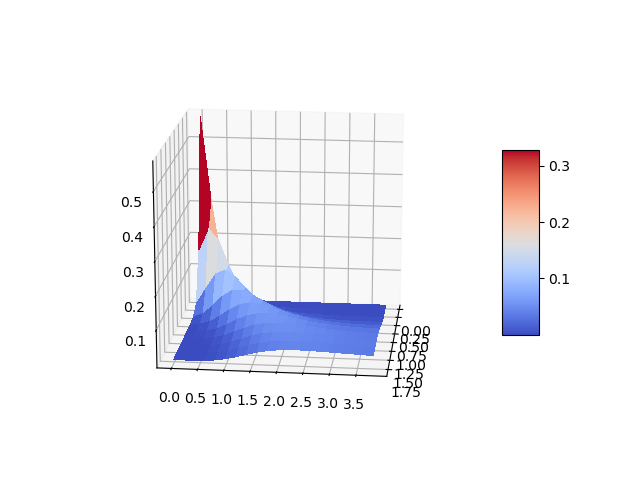}
        \caption{$\mu_X=3$, $\mu_Y=5$ for group $3$}
    \end{subfigure}%
    ~
    \newline
    \begin{subfigure}[t]{0.5\textwidth}
        \centering
        \includegraphics[height=2in]{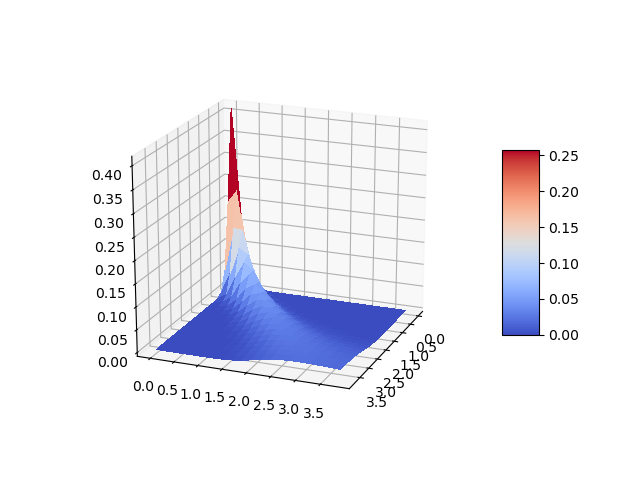}
        \caption{$\mu_X=7$, $\mu_Y=6$ group $4$}
    \end{subfigure}
~
    \begin{subfigure}[t]{0.5\textwidth}
        \centering
        \includegraphics[height=2in]{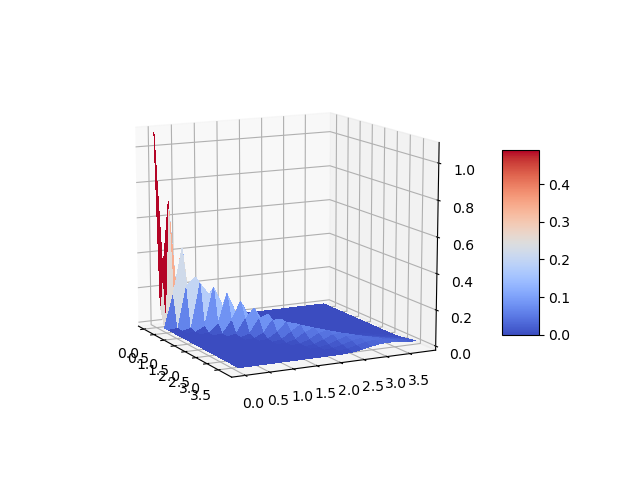}
        \caption{$\mu_X=7$, $\mu_Y=6$ for group $5$}
    \end{subfigure}
    \caption{Surface plots of density functions for \eqref{fco} based on different means and dependency groups.}\label{plotfco}
\end{figure}

In many studies with numerical data, the main questions are how to find the distribution of the data set, which can be univariate or multivariate, but the distribution of the existing data set is unknown in almost all research. So, they need a statistical estimator of distributions. The entropy concept is well known and used in many fields of study. The maximum entropy principle is a statistical method to find the best distribution dealing with inadequate information. Moreover, it acts acceptable with small sample sizes as well. Some intended constraints are required for maximum entropy methods based on such available information as moments. So, no strong assumptions are needed, which is another benefit of this method. While dealing with the univariate data set, it is easy to use the entropy procedure, and we are not worried about the loss of dependency between variables. In section \ref{EntropySec}, we found a joint distribution function according to some constraints whose result function is \eqref{JointME}, which is relatively dependence free. An important question is how intended conditions are defined to keep the original dependency between multivariate data sets. Which kinds of constraints guarantee the original dependency in the result distribution function? One way to reply to these questions is by using the copula functions. So, we find a copula function with the same dependency as the data set. To do this, we use the maximum entropy to get a copula function named the maximum copula entropy. We define some constraints under some dependency measures of data considering copula. Then, the result function has the same dependency on the available data set. The maximum copula function is transferred to the joint distribution function by the Sklar theorem. Thus, the unknown distribution of the practical set is obtained with the same dependency. \\
In this paper, we introduced a feasible way to get the distribution of an available data set by applying the maximum copula entropy. Afterwards, we get statistical control limits by exerting the joint density function, so the control limits are built with the original dependency between variables. Thus, the decision according to the limits is reliable. In the following, the power of the charts is exhibited by simulation study for different steps of shifts. Five groups of dependent measures are determined first in Table \ref{DG}, and all changes are applied for all groups. The scales are Spearman's rho, Blest $I$, and $III$ used in the number of conditions.
\begin{table}
 \caption{Triple measures for five dependency groups.}\label{DG}
 \centering
\scalebox{0.7}[0.7]{
\begin{tabular}{|c|c|c|c|}
\hline
Dependence group & $ \rho $ & $ \nu_1 $ & $ \eta $ \\ \hline
Group $ 1 $ & $ -0.4 $ & $ -0.5 $ & $ 0.2 $ \\ \hline
Group $ 2 $ & $ -0.1 $ & $ -0.18 $ & $ 0.45 $ \\ \hline
Group $ 3 $ & $ 0 $ & $ 0 $ & $ 0.5 $ \\ \hline
Group $ 4 $ & $ 0.1 $ & $ 0.18 $ & $ 0.55 $ \\ \hline
Group $ 5 $ & $ 0.4 $ & $ 0.5 $ & $ 0.8 $ \\ \hline
\end{tabular}
}
\end{table}\\
Coefficients of $ c(u,v) $ are calculated for dependence groups represented in Table \ref{coefco} as well as their corresponding surface plots in Figure \ref{plotmed}. The Lagrange coefficients are estimated according to the function \eqref{MaxCopulaE} with $r=5$. Various dependency values affect differently on the maximum copula function. We use those copula functions to get the joint density functions of some samples with different means. The surface plots of \eqref{fco} for several options are drawn in Figure \ref{plotfco}. Note that if function \eqref{JointME} is used, all surfaces are the same for different dependencies, but the maximum copula function \eqref{fco} is various for dependency groups. Several dependence measures have efficacy on the density function, so ignoring them leads to misunderstandings of production process features.\\
\begin{table}[!h]
\caption{Coefficients of the maximum copula entropy}\label{coefco}
 \centering
\scalebox{0.7}[0.7]{
\begin{tabular}{|c|c|c|c|c|c|}
\hline
Lagrange coefficients & Group $ 1 $ & Group $ 2 $ & Group $ 3 $ & Group $ 4 $ & Group $ 5 $\\ \hline
$\lambda_0 $ & $407.1356$ & $30.132266$ & $-3.1513221$ & $-3.846356$ & $-4.821063$\\ \hline
$\lambda_1 $ & $-2177.948$ & $-155.724111$ & $7.6161238$ & $8.759351$ & $5.748566$\\ \hline
$\lambda_2 $ & $4386.528$ & $296.108893$ & $172.0824301$ & $536.398532$ & $2708.479688$\\ \hline
$\lambda_3 $ & $-3672.8246$ & $-293.533359$ & $-358.672181$ & $-1043.037482$ & $-4764.824872$\\ \hline
$\lambda_4 $ & $429.5542$ & $149.434091$ & $178.0929512$ & $479.692338$ & $2069.490123$\\ \hline
$\lambda_5 $ & $973.9363$ & $-7.548779$ & $0.8739453$ & $19.045948$ & $20.27722$\\ \hline
$\lambda_6 $ & $6847.6116$ & $479.951059$ & $-397.0290744$ & $-1137.668711$ & $-5447.555543$\\ \hline
$\lambda_7 $ & $-6611.4438$ & $-441.819605$ & $396.8804704$ & $1104.027338$ & $4800.811622$\\ \hline
$\lambda_8 $ & $6394.9248$ & $407.919066$ & $-396.6800862$ & $-1071.220397$ & $-4230.97144$\\ \hline
\end{tabular}
}
\end{table}
\begin{table}[!h]
\caption{$ UCL $ with confidence level of $ 1-\alpha $ for some means and different measures of dependence whose copula coefficients are in Table \ref{coefco}. The first type of error is approximating $ 0.05 $ respectively to each case.}\label{UCLalpha05}
\centering
\scalebox{0.8}[0.8]{
\begin{tabular}{|c|c|c|c|c|c|c|} \hline
\multirow{2}{*}{Dependence groups}  & \multicolumn{2}{|c|}{${\mu }_X=2, {\mu }_Y=1$} & \multicolumn{2}{|c|}{${\mu }_X=3, {\mu }_Y=5$} & \multicolumn{2}{|c|}{${\mu }_X=7, {\mu }_Y=6$} \\ \cline{2-7}
 & $1-\alpha $ & $UCL$ & $1-\alpha $ & $UCL$ & $1-\alpha $ & $UCL$ \\ \hline
Group $ 1 $ & $0.950221$ & $2.57147$ & $0.95025$ & $2.55468$ & $0.950093$ & $2.66865$ \\ \hline
Group $ 2 $ & $0.950009$ & $6.65817$ & $0.950427$ & $2.55568$ & $0.950271$ & $6.76327$ \\ \hline
Group $ 3 $  & $0.950084$ & $8.51353$ & $0.950045$ & $8.50172$ & $0.950129$ & $8.49472$ \\ \hline
Group $ 4 $  & $0.950169$ & $8.6212$ & $0.950074$ & $8.59518$ & $0.950285$ & $8.60911$ \\ \hline
Group $ 5 $  & $0.950216$ & $8.11819$ & $0.950027$ & $8.04674$ & $0.950587$ & $8.17985$ \\ \hline
\end{tabular}
}
\end{table}
\begin{table}[!h]
 \caption{$ ARL_0 $ with $\alpha=0.05$ for different means.} \label{arl0}
 \centering
 \scalebox{0.73}[0.73]{
\begin{tabular}{|c|c|c|c|c|c|c|c|c|c|c|}
\hline
\multirow{3}{*}{Different means} & \multicolumn{10}{|c|}{Dependence group}  \\ \cline{2-11}
& \multicolumn{2}{|c|}{Group$1$}& \multicolumn{2}{|c|}{Group$2$}& \multicolumn{2}{|c|}{Group$3$}& \multicolumn{2}{|c|}{Group$4$}& \multicolumn{2}{|c|}{Group$5$} \\ \cline{2-11}
& Mean & variance &  Mean & variance &  Mean & variance &  Mean & variance &  Mean & variance \\ \hline
$\mu_X=2,~\mu_Y=1$ & $31.355$	& $819.878$	& $27.937$	& $710.053$	& $22.462$	& $464.647$	& $25.016$	& $528.356$	& $22.404$	& $418.417$ \\ \hline
$\mu_X=3,~\mu_Y=5$ & $31.075$	& $792.859$	& $8.928$	& $50.836$	& $23.537$	& $501.814$	& $25.157$	& $565.758$	& $22.044$	& $423.617$ \\ \hline
$\mu_X=7,~\mu_Y=6$ & $41.045$	& $1645.209$	& $27.209$	& $657.729$	& $23.060$	& $434.728$	& $25.161$	& $592.440$	& $23.198$	& $462.193$ \\ \hline
\end{tabular}
}
\end{table}
In Table \ref{UCLalpha05}, there are three means for $X$ and $Y$, which are $ \mu_X=2,~3,~7 $, and $ \mu_Y=1,~5,~6 $. These means are applied to find the marginal functions of $ X $ and $ Y $. They are estimated via a univariate maximum entropy method, as presented in section \ref{EntropySec}. The basic negligible first type of error is $0.05$. In this regard, the $UCL$s satisfies equation \eqref{InequUCL}. Tables \ref{arl0}, \ref{arl11}, \ref{arl12}, \ref{arl13} consist means and variance of $ARL_0$ and $ARL_1$. All $ARL$ is recalculated $1000$ times. According to $ARL_0$'s definition, the bigger $ARL_0$, the better performance of the control limits. The base value of $ARL_0$ for $\alpha=0.05$ is $20$. The efficiency of the maximum copula entropy is better than the basic in Table \ref{arl0}, where almost all of them are greater than $20$. Conversely of $ARL_0$, the smaller $ARL_1$, the superior implementation of the control chart. Some steps of shifts are needed to calculate $ARL_1$ that we make a mean model of changes for $X$ and $Y$ as:
\begin{displaymath}
\left\{\begin{array}{ll}
\mu_X '&=\mu_X+\delta_X \sigma_X, \\
\mu_Y '&=\mu_Y+\delta_Y \sigma_Y,
\end{array}\right.
\end{displaymath}
This paper aims to detect soft shifts that classical manners are unable to discover. The control chart is mighty because it is based on a fit distribution of data, but the traditional basic distribution is Normal and global for all process data. The $ARL_1$ method of generating is related to its definition. So, we do not use the $\beta$ in the calculations. In this regard, the corresponding $\beta$ is gotten by $ARL_1$'s values. For example, when $ARL_1=4.171$ for $\mu_X=2$ and $\mu_Y=1$ in group $1$, the $\beta$ is computed about $0.24$. This $\beta$ is affected by the $\alpha$ considering equation \eqref{InequUCL} applied in the $UCL$ calculations. $ARL_1$s are provided for different margins in Tables \ref{arl11}, \ref{arl12}, and \ref{arl13} for shift steps $0.1$, $0.5$, and $1$. Their mean and variance approximately obey the features of geometric distribution as well, and they are decreasing as the shifts are becoming larger. Since $ARL_1$s are so small with an error of $0.05$, they prevent wastage of capital and energy, and the accuracy of decisions is increased by using such control limits. Thus, the control limits of the maximum copula entropy have a preferable performance for undesirable soft shifts, which is difficult to detect for traditional methods. As a result, they are also suitable for observing larger changes. Thus, while using the control limit, we can ensure that a wide range of changes is easily detectable.
\begin{table}[!h]
 \caption{$ ARL_1 $ when $ \mu_X=2 $ and $ \mu_Y=1 $} \label{arl11}
 \centering
 \scalebox{0.7}[0.7]{
\begin{tabular}{|c|c|c|c|c|c|c|c|c|c|c|c|}
\hline
\multirow{3}{*}{$X$'s shifts} & \multirow{3}{*}{$Y$'s shifts} & \multicolumn{10}{|c|}{Dependence groups} \\ \cline{3-12}
 & & \multicolumn{2}{|c|}{Group $1$} & \multicolumn{2}{|c|}{Group $2$} & \multicolumn{2}{|c|}{Group $3$} & \multicolumn{2}{|c|}{Group $4$} & \multicolumn{2}{|c|}{Group $5$} \\ \cline{3-12}
 & & Mean & variance &  Mean & variance &  Mean & variance &  Mean & variance &  Mean & variance \\
\hline
\multirow{4}{*}{$ \delta_X=0.1 $} & $ \delta_Y=0 $ & $ 32.542 $ & $ 1022.793 $ & $ 28.012 $ & $ 698.124 $ & $ 22.836 $ & $ 431.109 $ & $ 24.663 $ & $ 472.750 $ & $ 21.795 $ & $ 383.91 $ \\ \cline{2-12}
&  $ \delta_Y=0.1 $ & $ 16.326 $ & $ 226.981 $ & $ 23.598 $ & $ 459.605 $ & $ 19.152 $ & $ 316.726 $ & $ 20.960 $ & $ 368.750 $ & $ 18.360 $ & $ 289.122 $ \\ \cline{2-12}
&  $ \delta_Y=0.5 $ & $ 4.565 $ & $ 9.099 $ & $ 12.278 $ & $ 111.551 $ & $ 11.118 $ & $ 95.357 $ & $ 11.327 $ & $ 98.064 $ & $ 10.817 $ & $ 85.865 $ \\ \cline{2-12}
&  $ \delta_Y=1 $ & $ 4.288 $ & $ 7.932 $ & $ 6.916 $ & $ 26.947 $ & $ 7.264 $ & $ 25.841 $ & $ 7.563 $ & $ 36.621 $ & $ 7.106 $ & $ 30.816 $ \\ \hline
\multirow{4}{*}{$ \delta_X=0.5 $} & $ \delta_Y=0 $ & $ 31.202 $ & $ 867.716 $ & $ 27.167 $ & $ 680.873 $ & $ 23.965 $ & $ 467.924 $ & $ 23.965 $ & $ 467.924 $ & $ 23.244 $ & $ 427.704 $ \\ \cline{2-12}
&  $ \delta_Y=0.1 $ & $ 23.244 $ & $ 427.704 $ & $ 24.085 $ & $ 487.198 $ & $ 18.944 $ & $ 268.881 $ & $ 20.635 $ & $ 380.476 $ & $ 19.137 $ & $ 301.848 $ \\ \cline{2-12}
&  $ \delta_Y=0.5 $ & $ 4.582 $ & $ 9.782 $ & $ 12.176 $ & $ 111.103 $ & $ 10.887 $ & $ 86.533 $ & $ 11.111 $ & $ 86.724 $ & $ 11.330 $ & $ 98.223 $ \\ \cline{2-12}
&  $ \delta_Y=1 $ & $ 4.211 $ & $ 8.073 $ & $ 6.856 $ & $ 25.671 $ & $ 7.340 $ & $ 32.595 $ & $ 7.251 $ & $ 31.274 $ & $ 7.138 $ & $ 32.067 $ \\ \hline
\multirow{4}{*}{$ \delta_X=1 $} & $ \delta_Y=0 $ & $ 32.854 $ & $ 850.620 $ & $ 27.681 $ & $ 699.654 $ & $ 22.845 $ & $ 441.568 $ & $ 23.85 $ & $ 520.022 $ & $ 22.324 $ & $ 434.010 $ \\ \cline{2-12}
&  $ \delta_Y=0.1 $ & $ 16.130 $ & $ 184.141 $ & $ 23.696 $ & $ 521.899 $ & $ 18.417 $ & $ 295.962 $ & $ 20.810 $ & $ 396.427 $ & $ 18.934 $ & $ 302.266 $ \\ \cline{2-12}
&  $ \delta_Y=0.5 $ & $ 4.401 $ & $ 7.451 $ & $ 11.906 $ & $ 111.007 $ & $ 10.994 $ & $ 83.304 $ & $ 10.751 $ & $ 78.850 $ & $ 10.914 $ & $ 88.744 $ \\ \cline{2-12}
&  $ \delta_Y=1 $ & $ 4.171 $ & $ 6.703 $ & $ 6.863 $ & $ 27.930 $ & $ 7.693 $ & $ 39.513 $ & $ 7.224 $ & $ 32.401 $ & $ 6.850 $ & $ 28.460 $ \\ \hline
\end{tabular}
}
\end{table}
\begin{table}[!h]
\caption{$ ARL_1 $ when $ \mu_X=3 $ and $ \mu_Y=5 $} \label{arl12}
\centering
\scalebox{0.7}[0.7]{
\begin{tabular}{|c|c|c|c|c|c|c|c|c|c|c|c|}
\hline
\multirow{3}{*}{$X$'s shifts} & \multirow{3}{*}{$Y$'s shifts} & \multicolumn{10}{|c|}{Dependence groups} \\ \cline{3-12}
 & & \multicolumn{2}{|c|}{Group $1$} & \multicolumn{2}{|c|}{Group $2$} & \multicolumn{2}{|c|}{Group $3$} & \multicolumn{2}{|c|}{Group $4$} & \multicolumn{2}{|c|}{Group $5$} \\ \cline{3-12}
 & & Mean & variance &  Mean & variance &  Mean & variance &  Mean & variance &  Mean & variance \\
\hline
\multirow{4}{*}{$ \delta_X=0.1 $} & $ \delta_Y=0 $ & $ 32.813 $ & $ 977.826 $ & $ 8.912 $ & $ 56.347 $ & $ 22.372 $ & $ 479.534 $ & $ 25.903 $ & $ 550.030 $ & $ 21.148 $ & $ 384.032 $ \\ \cline{2-12}
&  $ \delta_Y=0.1 $ & $ 15.458 $ & $ 185.348 $ & $ 7.441 $ & $ 35.481 $ & $ 19.821 $ & $ 343.291 $ & $ 19.595 $ & $ 327.403 $ & $ 18.157 $ & $ 249.111 $ \\ \cline{2-12}
&  $ \delta_Y=0.5 $ & $ 4.416 $ & $ 7.219 $ & $ 5.200 $ & $ 13.7923 $ & $ 11.301 $ & $ 91.144 $ & $ 10.959 $ & $ 91.580 $ & $ 11.442 $ & $ 92.635 $ \\ \cline{2-12}
&  $ \delta_Y=1 $ & $ 4.164 $ & $ 6.061 $ & $ 4.269 $ & $ 7.605 $ & $ 7.288 $ & $ 31.501 $ & $ 7.418 $ & $ 34.317 $ & $ 6.960 $ & $ 27.052 $ \\ \hline
\multirow{4}{*}{$ \delta_X=0.5 $} & $ \delta_Y=0 $ & $ 31.731 $ & $ 836.836 $ & $ 8.977 $ & $ 52.560 $ & $ 23.267 $ & $ 515.985 $ & $ 25.976 $ & $ 557.825 $ & $ 21.891 $ & $ 379.414 $ \\ \cline{2-12}
&  $ \delta_Y=0.1 $ & $ 16.190 $ & $ 203.555 $ & $ 7.829 $ & $ 37.334 $ & $ 19.505 $ & $ 324.951 $ & $ 20.671 $ & $ 355.171 $ & $ 18.489 $ & $ 270.234 $ \\ \cline{2-12}
&  $ \delta_Y=0.5 $ & $ 4.514 $ & $ 9.354 $ & $ 5.015 $ & $ 12.175 $ & $ 10.396 $ & $ 82.576 $ & $ 10.816 $ & $ 79.830 $ & $ 10.781 $ & $ 80.083 $ \\ \cline{2-12}
&  $ \delta_Y=1 $ & $ 4.100 $ & $ 6.369 $ & $ 4.158 $ & $ 6.638 $ & $ 7.233 $ & $ 34.486 $ & $ 7.053 $ & $ 28.445 $ & $ 6.770 $ & $ 27.408 $ \\ \hline
\multirow{4}{*}{$ \delta_X=1 $} & $ \delta_Y=0 $ & $ 33.194 $ & $ 1085.540 $ & $ 8.971 $ & $ 56.333 $ & $ 23.243 $ & $ 476.359 $ & $ 23.243 $ & $ 476.359 $ & $ 21.847 $ & $ 426.765 $ \\ \cline{2-12}
&  $ \delta_Y=0.1 $ & $ 15.947 $ & $ 216.730 $ & $ 7.653 $ & $ 34.365 $ & $ 18.494 $ & $ 257.132 $ & $ 21.060 $ & $ 405.754 $ & $ 20.335 $ & $ 334.980 $ \\ \cline{2-12}
&  $ \delta_Y=0.5 $ & $ 4.550 $ & $ 8.280 $ & $ 5.297 $ & $ 14.682 $ & $ 11.210 $ & $ 91.010 $ & $ 10.893 $ & $ 91.104 $ & $ 10.281 $ & $ 72.426 $ \\ \cline{2-12}
&  $ \delta_Y=1 $ & $ 4.121 $ & $ 6.275 $ & $ 4.124 $ & $ 7.301 $ & $ 7.340 $ & $ 30.844 $ & $ 7.247 $ & $ 31.471 $ & $ 6.742 $ & $ 29.842 $ \\ \hline
\end{tabular}
}
\end{table}
\begin{table}[h]
 \caption{$ ARL_1 $ when $ \mu_X=7 $ and $ \mu_Y=6 $} \label{arl13}
 \centering
 \scalebox{0.7}[0.7]{
\begin{tabular}{|c|c|c|c|c|c|c|c|c|c|c|c|}
\hline
\multirow{3}{*}{$X$'s shifts} & \multirow{3}{*}{$Y$'s shifts} & \multicolumn{10}{|c|}{Dependence groups} \\ \cline{3-12}
 & & \multicolumn{2}{|c|}{Group $1$} & \multicolumn{2}{|c|}{Group $2$} & \multicolumn{2}{|c|}{Group $3$} & \multicolumn{2}{|c|}{Group $4$} & \multicolumn{2}{|c|}{Group $5$} \\ \cline{3-12}
 & & Mean & variance &  Mean & variance &  Mean & variance &  Mean & variance &  Mean & variance \\
\hline
\multirow{4}{*}{$ \delta_X=0.1 $} & $ \delta_Y=0 $ & $ 37.914 $ & $ 1367.884 $ & $ 28.839 $ & $ 714.480 $ & $ 22.773 $ & $ 437.063 $ & $ 24.991 $ & $ 488.456 $ & $ 23.238 $ & $ 476.671 $ \\ \cline{2-12}
&  $ \delta_Y=0.1 $ & $ 20.635 $ & $ 394.610 $ & $ 24.542 $ & $ 523.843 $ & $ 19.251 $ & $ 333.922 $ & $ 20.044 $ & $ 327.204 $ & $ 19.099 $ & $ 288.448 $ \\ \cline{2-12}
&  $ \delta_Y=0.5 $ & $ 4.656 $ & $ 8.963 $ & $ 12.600 $ & $ 115.441 $ & $ 11.122 $ & $ 89.440 $ & $ 11.536 $ & $ 112.924 $ & $ 10.422 $ & $ 84.729 $ \\ \cline{2-12}
&  $ \delta_Y=1 $ & $ 4.165 $ & $ 6.893 $ & $ 6.761 $ & $ 29.426 $ & $ 7.545 $ & $ 31.372 $ & $ 7.305 $ & $ 31.392 $ & $ 7.022 $ & $ 33.216 $ \\ \hline
\multirow{4}{*}{$ \delta_X=0.5 $} & $ \delta_Y=0 $ & $ 38.526 $ & $ 1365.249 $ & $ 27.402 $ & $ 695.918 $ & $ 22.005 $ & $ 454.189 $ & $ 24.345 $ & $ 496.639 $ & $ 23.341 $ & $ 454.070 $ \\ \cline{2-12}
&  $ \delta_Y=0.1 $ & $ 20.390 $ & $ 341.434 $ & $ 24.840 $ & $ 528.936 $ & $ 24.840 $ & $ 528.936 $ & $ 20.150 $ & $ 337.094 $ & $ 19.938 $ & $ 337.481 $ \\ \cline{2-12}
&  $ \delta_Y=0.5 $ & $ 4.534 $ & $ 9.573 $ & $ 12.725 $ & $ 123.402 $ & $ 10.351 $ & $ 75.765 $ & $ 11.521 $ & $ 103.358 $ & $ 10.172 $ & $ 79.866 $ \\ \cline{2-12}
&  $ \delta_Y=1 $ & $ 4.246 $ & $ 7.057 $ & $ 7.219 $ & $ 29.654 $ & $ 7.163 $ & $ 31.682 $ & $ 7.427 $ & $ 35.455 $ & $ 6.958 $ & $ 28.58 $ \\ \hline
\multirow{4}{*}{$ \delta_X=1 $} & $ \delta_Y=0 $ & $ 40.114 $ & $ 1553.225 $ & $ 29.374 $ & $ 738.088 $ & $ 22.813 $ & $ 479.399 $ & $ 25.677 $ & $ 552.555 $ & $ 22.919 $ & $ 455.596 $ \\ \cline{2-12}
&  $ \delta_Y=0.1 $ & $ 21.049 $ & $ 400.818 $ & $ 25.885 $ & $ 572.166 $ & $ 18.691 $ & $ 281.885 $ & $ 20.329 $ & $ 310.796 $ & $ 18.500 $ & $ 294.849 $ \\ \cline{2-12}
&  $ \delta_Y=0.5 $ & $ 4.628 $ & $ 9.766 $ & $ 12.608 $ & $ 117.483 $ & $ 10.385 $ & $ 82.205 $ & $ 10.801 $ & $ 85.260 $ & $ 10.313 $ & $ 79.656 $ \\ \cline{2-12}
&  $ \delta_Y=1 $ & $ 4.201 $ & $ 7.920 $ & $ 6.739 $ & $ 31.503 $ & $ 7.371 $ & $ 36.881 $ & $ 7.040 $ & $ 28.521 $ & $ 6.801 $ & $ 27.961 $ \\ \hline
\end{tabular}
}
\end{table}

\section{Real data examples}\label{RealSec}
In this paper, we investigate a new method of finding statistical control limits. We estimate the dependence distribution via the maximum entropy principle and copula function. The copula function is used to preserve the main dependency in the database. A simulation section is added to show the performance of the presented method. Finally, we are going to peruse two areal data examples.
\subsection{A production process quality}\label{SecEx1}
\begin{figure}[t!]
    \centering
    \begin{subfigure}[t]{0.5\textwidth}
        \centering
        \includegraphics[height=2in]{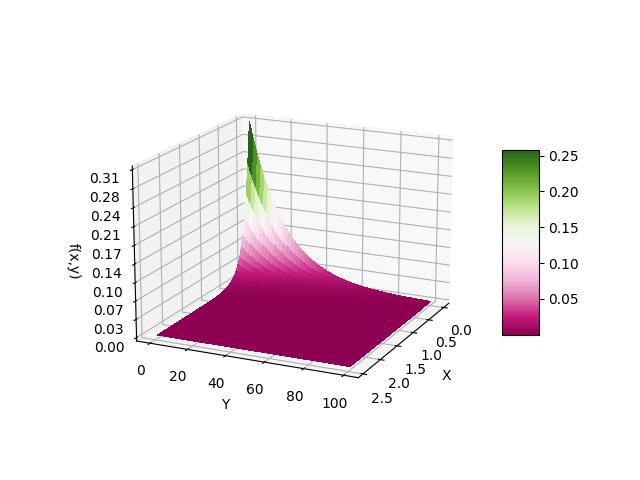}
        \caption{Three constraints based}\label{EX1Con3}
    \end{subfigure}%
    ~
    \centering
    \begin{subfigure}[t]{0.5\textwidth}
        \centering
        \includegraphics[height=2in]{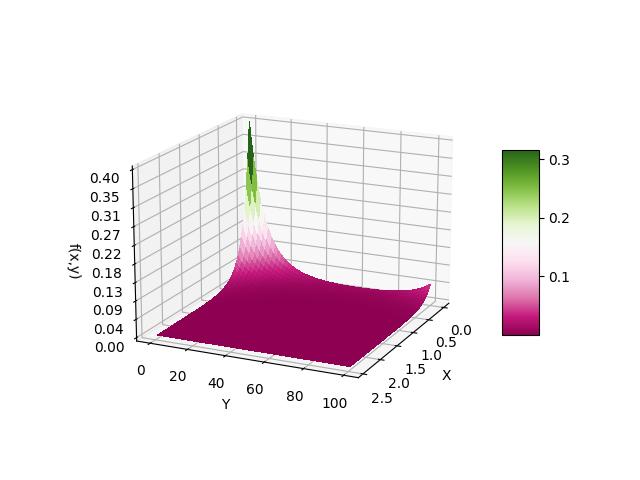}
        \caption{Five constraints based}\label{EX1Con5}
    \end{subfigure}%
    ~
    \newline
     \centering
    \begin{subfigure}[t]{0.5\textwidth}
        \centering
        \includegraphics[height=2in]{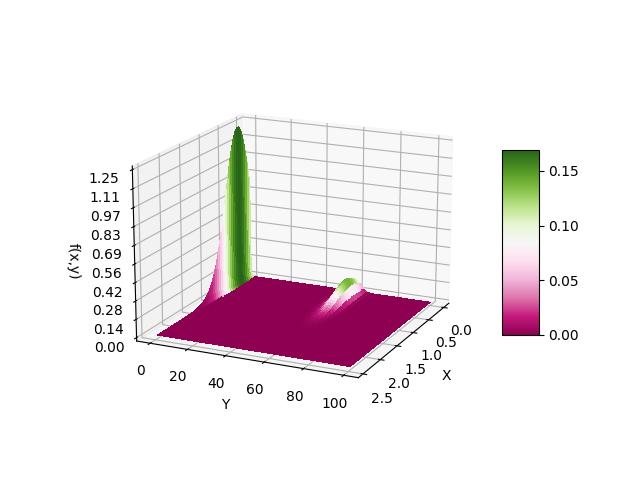}
        \caption{Seven constraints based}\label{EX1Con7}
    \end{subfigure}
    \caption{Surface plots of maximum entropy density functions for Table \ref{MaxEntEx1}}\label{MaxEnt1fig}
\end{figure}

\begin{figure}
  \centering
  \includegraphics[width=80mm]{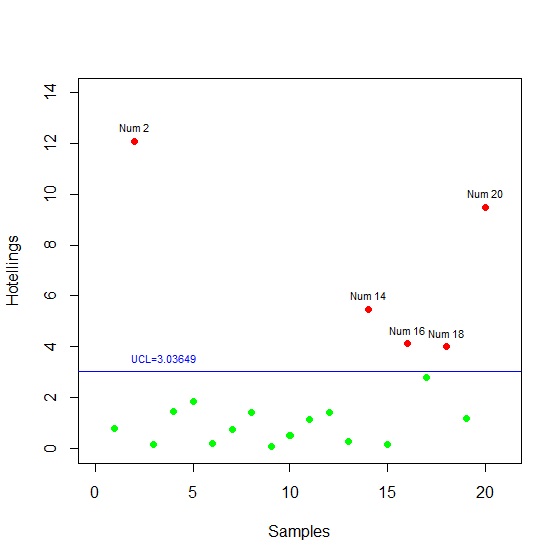}\\
  \caption{Control chart for the first $20$ samples in the first stage.}\label{Ex1F}
\end{figure}

\begin{figure}[t!]
    \centering
    \begin{subfigure}[t]{0.5\textwidth}
        \centering
        \includegraphics[height=2.5in]{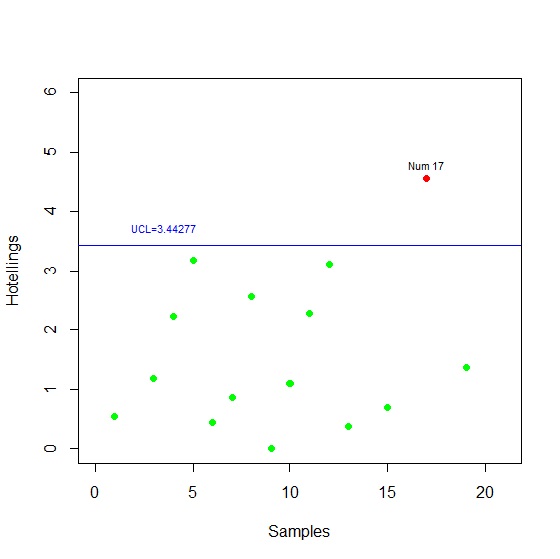}
        \caption{Sample $17$ is out-of-control.}
    \end{subfigure}%
    ~
    \centering
    \begin{subfigure}[t]{0.5\textwidth}
        \centering
        \includegraphics[height=2.5in]{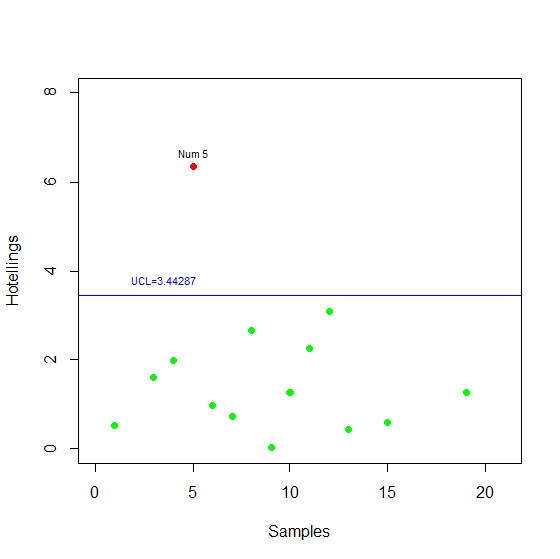}
        \caption{Sample $5$ is out-of-control.}
    \end{subfigure}%
    \caption{Control plots of stage $2$ and $3$. In each stage, samples $17$ and $5$ are removed in order to find a suitable $UCL$.}\label{Ex1P23}
\end{figure}

The first example is chosen from \cite{Quesenberry}, whose data includes eleven different quality variables from a production process. The data set has $30$ samples provided over time. In this regard, we are focusing on the first two quality characteristics as what paper \cite{Vargas} has done. The data with some notices are in Table \ref{Ex1}. The quality variables are nominated as $X$ and $Y$, respectively. Since the first phase of the data is required to obtain control limits, it is assumed that the first twenty samples belong to phase one. In this example, we discuss in detail how to get the $UCL$ via the paper method. The first step is to calculate the marginal distribution of the variables by their means:
\begin{displaymath}
\left\{\begin{array}{ll}
\int_{\mathbb{S}_X} d F_X(x)&=1, \\
\int_{\mathbb{S}_X} x~d F_X(x)&=0.53845, \\
\end{array}\right.
\end{displaymath}
and
\begin{displaymath}
\left\{\begin{array}{ll}
\int_{\mathbb{S}_Y} d F_Y(y)&=1, \\
\int_{\mathbb{S}_Y} y~d F_Y(y)&=59.9369. \\
\end{array}\right.
\end{displaymath}
The maximum Shannon entropy for margins are:
\begin{displaymath}
\left\{\begin{array}{ll}
  f_X(x) &= \exp (0.619060- 1.857182~x),~~ x\in \mathbb{S}_X, \\
  f_Y(y) &= \exp (-5.27750420 +0.01221784~y),~~ y\in \mathbb{S}_Y. \\
\end{array}\right.
\end{displaymath}
The next step is to figure out the joint density function based on the copula function. We calculate the dependencies via the estimators presented in section \ref{CopulaSec} or any other available estimators that are possible to use. Note that the presented estimators are sufficient when the sample size is large enough and depend on the data. The estimated dependency values are:
\begin{displaymath}
\left\{\begin{array}{ll}
\widehat{\rho}&=0.5636842,\\
\widehat{\eta}&=0.7218584,\\
\widehat{\nu_1}&=0.2685579,\\
\widehat{\nu_2}&=0.2972395.
\end{array}\right.
\end{displaymath}
Then, we have to make the constraints for calculation of maximum copula entropy:
\begin{equation}
 \begin{cases}
  \int \int_{I^2} c(u,v)dudv&=1, \\
  \int \int_{I^2} u^i c(u,v)dudv&=\frac{1}{i+1}, for ~i=1,\ldots,5, \\
  \int \int_{I^2} v^i c(u,v)dudv&=\frac{1}{i+1}, for ~i=1,\ldots,5, \\
  \int\int_{I^2} uv~ c(u,v)dudv&=\frac{\widehat{\rho}+3}{12}, \\
  \int\int_{I^2} u^2v~ c(u,v)dudv&=\frac{2\widehat{\rho}-\widehat{\nu_1}+2}{12}, \\
  \int\int_{I^2} uv^2~ c(u,v)dudv&=\frac{2\widehat{\rho}-\widehat{\nu_2}+2}{12}, \\
  \int\int_{I^2} u^2v^2~ c(u,v)dudv&=\frac{\widehat{\eta} + \frac{1}{5}}{6}.
  \end{cases} \label{measured}
\end{equation}
The second and third conditions, and also the fifth and sixth are merged to reduce the calculations as we explained them before. The maximum copula entropy concerning the Shannon is gotten:
\begin{align}\label{MaxCopuEntFunctionEx1}
c(u,v) = \exp & \left( -1021.614+4921.810~(u+v)-7979.657~(u^2+v^2) \right. \\ \nonumber
& +3137.956~(u^3+v^3)+4030.942~(u^4+v^4) \\ \nonumber
& -3786.658~(u^5+v^5)-15640.705~ uv \\ \nonumber
& \left. +14459.561 ~(u^2v+uv^2)-13364.703 ~u^2v^2\right),~\forall u,v\in [0,1].
\end{align}
Therefore, the joint distribution function is computed by \eqref{fco}, multiply $c(u,v)$  by the margins, and substituting $u$ and $v$ by $F_X(x)$ and $F_Y(y)$, respectively. Moreover, we calculate the maximum entropy function of the first $20$ samples with different numbers of constraints shown in Table \ref{MaxEntEx1}. The joint density is so simpler than the maximum copula entropy. We see in Table \ref{MaxEntEx1} that the function is almost constant when the number of constraints is $9$ and $11$. So, the function is an overestimation in these cases. Therefore, we choose one of the density functions with $3$, $5$, and $7$ constraints. In this regard, we plot them in Figure \ref{MaxEnt1fig} to decide which of them is a proper choice. The density function is underestimation in Figure \ref{EX1Con3}. Also, it seems to be an incomplete plot on the tail in Figure \ref{EX1Con5}. The maximum entropy approximates the joint density function based on $7$-constraint well shown in Figure \ref{EX1Con7}. Thus, the joint density function-based maximum entropy is:
\begin{align}\label{MaxEntFunctionEx1}
f_{X,Y}(x,y)=\exp &\left( -0.26667567+3.03875779x-4.03869815y\right.\\ \nonumber
&-4.83540788x^2+0.13326476y^2\\ \nonumber
&\left.+0.44454878x^3-1.10889612e^{-03}y^3 \right), (x,y)\in \mathbb{S}(X,Y).
\end{align}
\begin{table}[!h]
\caption{Coefficients of the maximum entropy for the first example}\label{MaxEntEx1}
 \centering
\scalebox{0.7}[0.7]{
\begin{tabular}{|c|c|c|c|c|c|}
\hline
\multirow{2}{*}{Lagrange coefficients of} & \multicolumn{5}{|c|}{Number of constraints} \\ \cline{2-6}
& $3$ & $5$ & $7$ & $9$ & $11$ \\ \hline
constant & $1.16599193$ & $9.19634366e^{-01}$ & $0.26667567$ & $6.90774436$ & $6.90775466$\\ \hline
$x $     & $5.95989323$ & $5.08152374$ & $-3.03875779$ & $3.61656946e^{-07}$ & $3.88173302e^{-09}$\\ \hline
$y $     & $0.05199653$ & $1.17971216e^{-01}$ & $4.03869815$ & $2.16995601e^{-08}$ & $2.26435444e^{-10}$\\ \hline
$x^2 $   & $0$ & $-4.19700033e^{-01}$ & $4.83540788$ & $5.42485627e^{-08}$ & $5.82260185e^{-10}$\\ \hline
$y^2 $   & $0$ & $-9.64389092e^{-04}$ & $-1.33264760e^{-01}$ & $2.53176986e^{-10}$ & $2.58799304e^{-12}$\\ \hline
$x^3 $   & $0$ & $0$ & $-4.44548777e^{-01}$ & $7.23316916e^{-09}$ & $7.76349871e^{-11}$\\ \hline
$y^3 $   & $0$ & $0$ & $1.10889612e^{-03}$ & $2.90899219e^{-12}$ & $2.92761868e^{-14}$\\ \hline
$x^4 $   & $0$ & $0$ & $0$ & $9.04179709e^{-10}$ & $9.70473376e^{-12}$\\ \hline
$y^4 $   & $0$ & $0$ & $0$ & $5.14080999e^{-14}$ & $5.18651582e^{-16}$\\ \hline
$x^5 $   & $0$ & $0$ & $0$ & $0$ & $1.16499313e^{-12}$\\ \hline
$y^5 $   & $0$ & $0$ & $0$ & $0$ & $2.90000001e^{-16}$\\ \hline
\end{tabular}
}
\end{table}
The final step is to get the $UCL$, and \eqref{InequUCL} have to be solved for the $UCL$. The $UCL$ is $3.03649$ for the copula-based density \eqref{MaxCopuEntFunctionEx1} and is $7.716048$ for the maximum entropy function \eqref{MaxEntFunctionEx1} at confidence level $\% 95$. It means that the $UCL$ with a lack of data dependency is less sensitive than the control limits, which meats the dependency. Thus, we use $UCL=3.03649$ related to the maximum copula entropy. \\
Montgomery \cite{Montgomery} explained that after calculating quality control limits, all sample from phase one has to be plotted to ensure that all samples are under control. Then, the control limits are reliable. On other hand, if one or more point is out-of-control, they have to be removed from data-based. This process continues until all members of the sample are within control. In Table \ref{Ex1}, four stages are presented to find $UCL$, which keeps all samples in-control. We draw the first $20$ samples along with the $UCL$ in Figure \ref{Ex1F}.
Five points are above the control limit in the first stage in Table \ref{Ex1}. So, we have to build another $UCL$ in the lake of those samples which are not in-control. The $T^2$-Hotelling values for the second phase are calculated based on the first $20$ samples in stage $1$. Bold values of samples are out-of-control, but this result is not reliable. The reason is that all points of phase one are not in-control in the first stage. We have to remove those points and build another $UCL$ as we do in the second stage shown in Table \ref{Ex1}. Thus, we recalculate the $UCL$ in stage $2$ with the remaining $15$ points. To do this, we must do all the calculations from the beginning. So, the marginal distributions are:
\begin{displaymath}
\left\{\begin{array}{ll}
  f_X(x) &= \exp (1.605991-1.833068~x),~~ x\in \mathbb{S}_X, \\
  f_Y(y) &= \exp (-4.29706292+0.01254347~y),~~ y\in \mathbb{S}_Y. \\
\end{array}\right.
\end{displaymath}
The dependency measures are:
\begin{displaymath}
\left\{\begin{array}{ll}
\widehat{\rho}&=0.6349206,\\
\widehat{\eta}&=0.9525656,\\
\widehat{\nu_1}&= 0.3331598,\\
\widehat{\nu_2}&=0.1635767.
\end{array}\right.
\end{displaymath}
So, the maximum copula entropy is:
\begin{align*}
c(u,v) = \exp &\left( -1022.656+4126.525~(u+v)-3264.327~(u^2+v^2) \right. \\ \nonumber
& -6298.043~(u^3+v^3)+11897.127~(u^4+v^4) \\ \nonumber
& -6017.781~(u^5+v^5)-14318.753~ uv  \\ \nonumber
& \left. +12658.220 ~(u^2v+uv^2)-11146.407 ~u^2v^2\right),~\forall u,v\in [0,1].
\end{align*}
The second $UCL$ is $3.44277$, and the only point, sample $17$, is out-of-control. The process of calculation is continued to the third step, where the $UCL$ is $3.44287$. This control chart calculation process is in Figure \ref{Ex1P23} for the second and third stages. We have to perform these preliminary steps on phase one when we are not sure whether the data of phase one are under control or not. We achieve a data set by filtration, which has the desired quality and controlled conditions. In the fourth step, we get the control limit that includes all the in-control samples of phase one. So, this limit of control will be reliable, and we are sure about its fine quality. The final $UCL$ is $2.87983$. The second part of the sample in Table \ref{Ex1} shows different values of $T^2$-Hotelling statistics for the rest of the data in the second phase. The bold values are out-of-control samples in each stage. We conclude that the control limits of the lower stage are not accurate enough to detect all unwanted changes in the process. So, we upgrade our determined control limit and increase the sensitivity of the $UCL$ by removing out-of-control samples from phase one.
\begin{table}[!h]
\caption{The sample along with the corresponding $T^2$-Hotelling values of the four different stages are presented to get the reliable control limits and detect the samples that have the undesired quality.} \label{Ex1}
\centering
\scalebox{0.7}[0.7]{
\begin{tabular}{|c|c|c|c|c|c|c|}
\hline
\multirow{2}{*}{Num} & \multirow{2}{*}{$X$} & \multirow{2}{*}{$Y$}	& \multicolumn{4}{|c|}{$T^2$-Hotelling}  \\ \cline{4-7}
& & & Stage $1$	& Stage $2$ &	Stage $3$	&  Stage $4$ \\ \hline
\multicolumn{7}{|c|}{Phase One} \\ \hline
$1$ & $	0.567$ & $	60.558	$ & $0.77889186$ & $	0.535972443	$ & $0.5215308	$ & $0.4509497$ \\ \hline
$2$ & $	0.538$ & $	56.303$ & \boldmath$	12.10300519	$ & $-$ & $-$ & $-$ \\ \hline
$3$ & $	0.53$ & $	59.524$ & $	0.18588867$ & $	1.177161371$ & $	1.59655517	$ & $2.7030141$ \\ \hline
$4$ & $	0.562$ & $	61.102$ & $	1.47298941$ & $	2.229342412	$ & $1.98567071$ & $	1.9318334$ \\ \hline
$5$ & $	0.483$ & $	59.834$ & $	1.85653453	$ & $3.183258805$ & \boldmath$	6.34453123$ & $-$ \\ \hline
$6$ & $	0.525	$ & $60.228	$ & $0.20325866	$ & $0.435849055$ & $	0.96401561	$ & $1.8910021$ \\ \hline
$7$ & $	0.556$ & $	60.756$ & $	0.74539794	$ & $0.859362303	$ & $0.71568288	$ & $0.6518001$ \\ \hline
$8$ & $	0.586$ & $	59.823	$ & $1.40816558	$ & $2.561668141$ & $	2.64819189	$ & $2.3410827$ \\ \hline
$9$ & $	0.547	$ & $60.153$ & $0.07992998$ & $	0.008095458	$ & $0.03036386	$ & $0.1891093$ \\ \hline
$10	$ & $0.531	$ & $60.64	$ & $0.50788714	$ & $1.096696583$ & $	1.2515963	$ & $1.6608542$ \\ \hline
$11	$ & $0.581	$ & $59.785	$ & $1.14767429$ & $	2.274062228$ & $	2.24026394$ & $	1.8909142$ \\ \hline
$12$ & $	0.585	$ & $59.675	$ & $1.4288771	$ & $3.111496984	$ & $3.07034593	$ & $2.6533813$ \\ \hline
$13$ & $	0.54	$ & $60.489	$ & $0.27757934$ & $	0.375753331$ & $	0.42863855$ & $	0.6615801$ \\ \hline
$14	$ & $0.458	$ & $61.067	$ & \boldmath$5.46768391$ & $-$ & $-$ & $-$ \\ \hline
$15	$ & $0.554	$ & $59.788	$ & $0.17655415	$ & $0.702630646	$ & $0.59595894$ & $	0.6095634$ \\ \hline
$16	$ & $0.469	$ & $58.64	$ & \boldmath$4.11715192$ & $-$ & $-$ & $-$ \\ \hline
$17$ & $	0.471$ & $	59.574	$ & $2.78971477	$ & \boldmath$4.558227247	$ & $-$ & $-$ \\ \hline
$18	$ & $0.457$ & $	59.718$ & \boldmath$	4.00679641$ & $-$ & $-$ & $-$ \\ \hline
$19$ & $	0.565	$ & $60.901	$ & $1.17923852$ & $	1.366716798$ & $	1.25308081$ & $	1.2191211$ \\ \hline
$20	$ & $0.664	$ & $60.18	$ & \boldmath$9.51709933$ & $-$ & $-$ & $-$ \\ \hline
\multicolumn{7}{|c|}{Phase Two} \\ \hline
$21$ & $	0.6	$ & $60.493	$ & \boldmath$4.9020358	$ & \boldmath $4.8327793$ & \boldmath $	6.64730669$ & \boldmath $	7.7459508$ \\ \hline
$22	$ & $0.586	$ & $58.37$ & \boldmath $	7.8444876$ & \boldmath $	29.5571898$ & \boldmath $	26.27275769	$ & \boldmath $23.3134032$ \\ \hline
$23$ & $	0.567	$ & $60.216	$ & $1.0707439	$ & $0.827601	$ & $0.77040238	$ & $0.4848948$ \\ \hline
$24$ & $	0.496$ & $	60.214	$ & $2.425637$ & \boldmath $	4.7412564	$ & \boldmath $8.65226246$ & \boldmath $	14.1333281$ \\ \hline
$25	$ & $0.485	$ & $59.5	$ & \boldmath $3.6386316	$ & \boldmath $6.4763706	$ & \boldmath $12.80958642	$ & \boldmath $22.3899082$ \\ \hline
$26	$ & $0.573$ & $	60.052$ & $	1.4442962$ & $	1.8457366$ & \boldmath $	1.82042288$ & \boldmath $	1.3979842$ \\ \hline
$27	$ & $0.52	$ & $59.501	$ & $0.6985124$ & $	2.7355863$ & $	4.4371601	$ & $7.9458876$ \\ \hline
$28	$ & $0.556$ & $	58.476$ & \boldmath $	4.4925447$ & \boldmath $	19.8322597$ & \boldmath $	17.52640091	$ & \boldmath $17.1415584$ \\ \hline
$29$ & $	0.539$ & $	58.666	$ & $2.9682209	$ & \boldmath $13.6351807	$ & \boldmath $12.85114487$ & \boldmath $	14.5623441$ \\ \hline
$30$ & $	0.554$ & $	60.239	$ & $0.4240915$ & $	0.1165282	$ & $0.02768135$ & $	0.0195593$ \\ \hline
\end{tabular}
}
\end{table}
\subsection{A flood events}
The second example is from Yue \cite{Yue}, where there are three variables, duration, volume, and peak, for the flood from $1919$ to $1995$. We select this example to say that the new control limit introduced in this paper is not proposed just for manufacturing processes, but also we use it to check the weather conditions in case of a storm. So, the specialists can use this control chart in research based on their needs.
The flood data was recorded in the Madawaska river basin in Quebec province, Canada. Cheng and Mukherjee \cite{Cheng} made a $T^2$-Hotelling control chart for the first two variables. They assumed the first $70$ samples belong to phase one, and the rest is in phase two. In this subsection, we make a proper control limit for the data according to the durations and volumes of floods. The variables are offered as $X$ and $Y$. The computation steps are briefly discussed here. As we mentioned in \ref{SecEx1}, the first step of the calculation is to get the marginal distribution for phase one of the data. The margins are gotten according to the maximum entropy principle concerning Shannon entropy based on the means. Those are  $80.25714$ and $9084.657$ for $X$ and $Y$, respectively. So, the density functions are:
\begin{displaymath}
\left\{\begin{array}{ll}
  f_X(x) &= \exp (-4.4864427-0.0108226~x),~~ x\in \mathbb{S}_X, \\
  f_Y(y) &= \exp (-9.640157-2.759943e^{-05}~y),~~ y\in \mathbb{S}_Y. \\
\end{array}\right.
\end{displaymath}
In this example, we assume that the dependencies between the variables have not changed over $110$ years, and the dependency measures over these years are:
\begin{displaymath}
\left\{\begin{array}{ll}
\widehat{\rho}&=0.4722444,\\
\widehat{\eta}&=0.860833,\\
\widehat{\nu_1}&=0.456628,\\
\widehat{\nu_2}&=0.4514329,\\
\end{array}\right.
\end{displaymath}
The copula function according to these measures and the presented conditions in \ref{SecEx1} are:
\begin{align*}
c(u,v) = \exp &\left( 6.580167-13.431977~(u+v)-11181.767618~(u^2+v^2) \right. \\ \nonumber
& +21517.583701~(u^3+v^3)-10246.181394~(u^4+v^4) \\ \nonumber
& -91.652515~(u^5+v^5)+22505.173734~ uv \\ \nonumber
& \left. -21703.580642 ~(u^2v+uv^2)+20931.154327 ~u^2v^2\right),~\forall u,v\in [0,1].
\end{align*}
So, the joint density function of $X$ and $Y$ is gotten by substituting the marginal and copula functions inside \eqref{fco}. Also, the corresponding Lagrange coefficients for the maximum entropy in the absence of dependency are shown in Table \ref{MaxEntEx2}. It seems that the maximum entropy is unable to approximate the joint density function well in this example because almost all coefficients are zero except the constants. In this case, the maximum entropy principle is useless in the bivariate mood in the lack of copula function.\\
\begin{table}[!h]
\caption{Coefficients of the maximum entropy for the seecond example}\label{MaxEntEx2}
 \centering
\scalebox{0.7}[0.7]{
\begin{tabular}{|c|c|c|c|c|c|}
\hline
\multirow{2}{*}{Lagrange coefficients of} & \multicolumn{5}{|c|}{Number of constraints} \\ \cline{2-6}
& $3$ & $5$ & $7$ & $9$ & $11$ \\ \hline
constant & $6.90775512$ & $6.90775528$ & $6.90775528$ & $6.90775528$ & $6.90775528$\\ \hline
$x $     & $2.82527006e^{-08}$ & $1.87317075e^{-12}$ & $1.27453129e^{-16}$ & $8.62325374e^{-21}$ & $5.78465447e^{-25}$\\ \hline
$y $     & $2.64918639e^{-10}$ & $1.56112298e^{-14}$ & $9.95825505e^{-19}$ & $6.46809475e^{-23}$ & $4.21841301e^{-27}$\\ \hline
$x^2 $   & $0$ & $1.40512786e^{-14}$ & $9.56068455e^{-19}$ & $6.46859039e^{-23}$ & $4.33926235e^{-27}$\\ \hline
$y^2 $   & $0$ & $3.35779397e^{-18}$ & $2.16519544e^{-22}$ & $1.42162559e^{-26}$ & $9.36199895e^{-31}$\\ \hline
$x^3 $   & $0$ & $0$ & $6.59103562e^{-21}$ & $4.45937835e^{-25}$ & $2.99144194e^{-29}$\\ \hline
$y^3 $   & $0$ & $0$ & $1.89919868e^{-22}$ & $1.28496368e^{-26}$ & $8.61979831e^{-31}$\\ \hline
$x^4 $   & $0$ & $0$ & $0$ & $2.17860956e^{-26}$ & $1.46145572e^{-30}$\\ \hline
$y^4 $   & $0$ & $0$ & $0$ & $1.28496366e^{-26}$ & $8.61979831e^{-31}$\\ \hline
$x^5 $   & $0$ & $0$ & $0$ & $0$ & $1.03697050e^{-30}$\\ \hline
$y^5 $   & $0$ & $0$ & $0$ & $0$ & $8.61979831e^{-31}$\\ \hline
\end{tabular}
}
\end{table}
The upper control limit is calculated by solving equation \eqref{InequUCL} concerning $UCL$, which is $6.85875$. Figure \ref{EX2FigStage1} is related to this part of the calculations. Four points of the sample are out-of-control, whose corresponding data is in Table \ref{Ex2}. So, we have to remove them and repeat the computations for the second stage. In this level, the four points are omitted from data-based, and new marginal distributions are:
\begin{displaymath}
\left\{\begin{array}{ll}
  f_X(x) &= \exp (-3.41630047-0.01171986~x),~~ x\in \mathbb{S}_X, \\
  f_Y(y) &= \exp (-8.597681-3.231503e^{-05}~y),~~ y\in \mathbb{S}_Y. \\
\end{array}\right.
\end{displaymath}
Since no change in the dependence between the two variables is assumed, the same copula function is used for this calculation step. The final $UCL$ is $6.89478$. All $66$ points of the data set are in the new statistical control limit shown in Figure \ref{EX2FigStage2}. The $T^2$-Hotelling values of phase two for two different stage is exhibited in Table \ref{Ex2} as well as the Hotelling statistics for phase one. The bold values are detected as out-of-control according to each step. There is an interesting point in the results. Sample numbers $81$ and $84$ in the first stage are known as out-of-control samples, while in the second stage they are within the control range. Comparing the duration and volumes of the samples along with the other in-control data, we realize that the detection in the second stage is nearly correct.
\begin{figure}[t!]
    \centering
    \begin{subfigure}[t]{0.5\textwidth}
        \centering
        \includegraphics[height=2.5in]{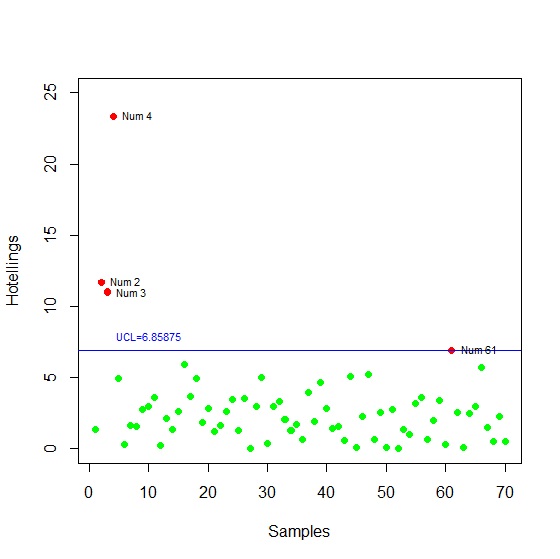}
        \caption{The out-of-control samples are detected.}\label{EX2FigStage1}
    \end{subfigure}%
    ~
    \centering
    \begin{subfigure}[t]{0.5\textwidth}
        \centering
        \includegraphics[height=2.5in]{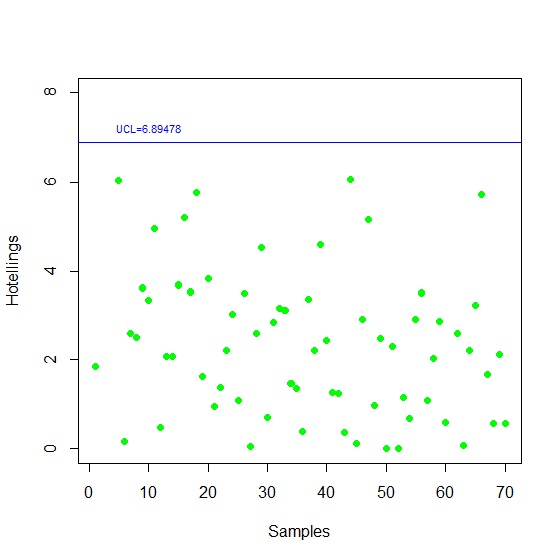}
        \caption{New control limit is defined according to the rest of samples.}\label{EX2FigStage1}
    \end{subfigure}%
    \caption{The second example control plots. In stage $2$, samples $2$, $3$, $4$, and $61$ are removed in order to get a appropriate $UCL$.}\label{Ex2P23}
\end{figure}
\begin{table}[!h]
\caption{The flood duration and volume are represented as $X$ and $Y $, respectively. Two columns belong to $T^2$-Hotelling values for two-stage of calculations. The first $70$ samples belong to phase one, and the rest of them were getting from phase two.The bold numbers are out-of-control.} \label{Ex2}
\centering
\scalebox{0.7}[0.7]{
\begin{tabular}{|c|c|c|c|c||c|c|c|c|c|}
\hline
\multirow{2}{*}{Num} & \multirow{2}{*}{$X$} & \multirow{2}{*}{$Y$}	& \multicolumn{2}{|c|}{$T^2$-Hotelling} & \multirow{2}{*}{Num} & \multirow{2}{*}{$X$} & \multirow{2}{*}{$Y$}	& \multicolumn{2}{|c|}{$T^2$-Hotelling} \\ \cline{4-5} \cline{9-10}
& & & Step $1$ & Step $2$ & & & & Stage $1$ & Stage $2$ \\ \hline
\multicolumn{5}{|c|}{Phase One}& $	56	$ & $	90	$ & $	13608	$ & $	3.55987914	$ & $	3.49396394$ \\ \hline
$1	$ & $100	$ & $	12057	$ & $	1.3554222	$ & $	1.84734622	$ & $57	$ & $93$ & $8949	$ & $0.63330617	$ & $1.07191071$ \\ \hline
$2	$ & $146	$ & $10853	$ & \boldmath$11.73821612$ & $-$ & $58	$ & $90		$ & $12577	$ & $1.98199188	$ & $	2.03091817$ \\ \hline
$3	$ & $142	$ & $10299	$ & \boldmath$11.02731534$ & $-$ & $59	$ & $65		$ & $11437		$ & $3.37692511	$ & $	2.8507401$ \\ \hline
$4	$ & $170	$ & $10818		\boldmath$ & $23.37140017$ & $-$ & $60	$ & $90		$ & $9266	$ & $0.2772227	$ & $0.59872804$ \\ \hline
$5		$ & $110	$ & $7748	$ & $4.91697816		$ & $6.046848	$ & $61	$ & $135	$ & $14559	$ & \boldmath$6.8977365$ & $-$  \\ \hline
$6	$ & $76$ & $9763	$ & $0.27308055	$ & $	0.16149275	$ & $62	$ & $68	$ & $5057	$ & $2.57471547	$ & $	2.58695985$ \\ \hline
$7	$ & $108	$ & $11127	$ & $1.66469209	$ & $2.57703041	$ & $63	$ & $80	$ & $9645	$ & $0.07876282	$ & $0.07882334$ \\ \hline
$8	$ & $107	$ & $10659	$ & $1.59281071		$ & $2.48771464	$ & $64	$ & $47	$ & $7241	$ & $2.49600239	$ & $2.20343276$ \\ \hline
$9	$ & $104	$ & $8327	$ & $2.7416352	$ & $3.6108401	$ & $65	$ & $97	$ & $13543	$ & $2.99103845	$ & $3.21294005$ \\ \hline
$10		$ & $100	$ & $13593	$ & $2.97106882	$ & $3.31707484	$ & $66	$ & $97	$ & $	15003	$ & $5.67049529	$ & $5.71978743$ \\ \hline
$11	$ & $120	$ & $12882	$ & $3.56245366	$ & $4.9677068	$ & $67	$ & $79	$ & $6460	$ & $1.5001489	$ & $1.65859997$ \\ \hline
$12	$ & $90	$ & $9957	$ & $0.20956068	$ & $0.48629727	$ & $68	$ & $78	$ & $7502	$ & $0.47740872	$ & $0.56736257$ \\ \hline
$13	$ & $60$ & $	5236	$ & $2.14394992	$ & $2.06127483	$ & $69	$ & $50		$ & $5650	$ & $2.28777597		$ & $2.12465613$ \\ \hline
$14	$ & $102	$ & $9581	$ & $1.3295513	$ & $2.06938059		$ & $70	$ & $76	$ & $7350	$ & $0.50615828	$ & $0.56009575$ \\ \hline
$15	$ & $	113$ & $	12740	$ & $2.64425299	$ & $3.68671936	$ & \multicolumn{5}{|c|}{Phase Two} \\ \hline
$16		$ & $52		$ & $11174	$ & $5.90031359	$ & $	5.20634663	$ & $71	$ & $101	$ & $9506	$ & $2.4747948	$ & $3.8821229$ \\ \hline
$17		$ & $42	$ & $4780	$ & $3.63192971	$ & $3.52337361	$ & $72	$ & $78	$ & $6728	$ & $2.2576101	$ & $2.5098685$ \\ \hline
$18	$ & $113	$ & $14890	$ & $4.91350588	$ & $5.75903123	$ & $73	$ & $73	$ & $13315	$ & \boldmath$10.9100293	$ & \boldmath$9.7480815$ \\ \hline
$19	$ & $52	$ & $6334	$ & $1.81490223	$ & $1.61452889	$ & $	74	$ & $71	$ & $8041	$ & $0.4259598	$ & $0.2490088$ \\ \hline
$20	$ & $109	$ & $9177	$ & $2.79752184	$ & $3.84906311	$ & $75	$ & $105	$ & $10174	$ & $2.9319441	$ & $4.600151$ \\ \hline
$21	$ & $57	$ & $7133	$ & $1.17824544		$ & $0.95635856	$ & $76	$ & $128	$ & $14769	$ & \boldmath$11.8738229	$ & \boldmath$15.3394817$ \\ \hline
$22	$ & $53	$ & $6865	$ & $1.61225285	$ & $1.3791327	$ & $77	$ & $65	$ & $8711	$ & $1.2905331	$ & $0.799728$ \\ \hline
$23	$ & $52	$ & $8918	$ & $2.6325595	$ & $2.20768402	$ & $78	$ & $85.346	$ & $10702.4	$ & $0.8234114	$ & $0.9608757$ \\ \hline
$24	$ & $47	$ & $8704	$ & $3.45462932	$ & $3.01879603	$ & $79	$ & $78.307	$ & $15229.7	$ & \boldmath$18.5380669	$ & \boldmath$17.1361787$ \\ \hline
$25	$ & $56$ & $	6907	$ & $1.299942	$ & $1.08578937	$ & $80	$ & $71.4491	$ & $14412.2	$ & \boldmath$17.1369061	$ & \boldmath$15.5733459$ \\ \hline
$26	$ & $60	$ & $4189	$ & $3.53565292	$ & $3.49237819	$ & $81	$ & $44.4383	$ & $8200.7	$ & \boldmath$7.1013877	$ & $6.3162928$ \\ \hline
$27	$ & $78	$ & $8637	$ & $0.02897507	$ & $0.05137817	$ & $82	$ & $75.102	$ & $13458.2	$ & \boldmath$10.7274652	$ & \boldmath$9.6434389$ \\ \hline
$28	$ & $48	$ & $8409	$ & $2.97549643	$ & $2.57583044	$ & $83	$ & $42.2206	$ & $9153.79	$ & \boldmath$10.3079061	$ & \boldmath$9.2970714$ \\ \hline
$29	$ & $79	$ & $13602	$ & $4.97916313	$ & $4.53415961	$ & $84	$ & $61.561	$ & $11364	$ & \boldmath$7.8878168	$ & $6.7256556$ \\ \hline
$30	$ & $89	$ & $8788	$ & $0.37958655	$ & $0.69686466	$ & $85	$ & $96.7068	$ & $18433.2	$ & \boldmath$31.9040035	$ & \boldmath$30.9385949$ \\ \hline
$31	$ & $47	$ & $5002	$ & $2.96758996	$ & $2.83879262	$ & $86	$ & $66.3522	$ & $15959.6	$ & \boldmath$30.2200327	$ & \boldmath$27.8671644$ \\ \hline
$32	$ & $43	$ & $5167	$ & $3.28535882	$ & $3.14969028	$ & $87	$ & $57.3523	$ & $13795.7	$ & \boldmath$21.6796619	$ & \boldmath$19.6697575$ \\ \hline
$33	$ & $109	$ & $10128	$ & $2.08397791	$ & $3.09652757	$ & $88	$ & $130.033	$ & $20245.3	$ & \boldmath$36.3264075	$ & \boldmath$38.7449082$ \\ \hline
$34	$ & $92	$ & $12035	$ & $1.29362259	$ & $1.4703014	$ & $89	$ & $81.2941	$ & $16645.3	$ & \boldmath$26.1934871	$ & \boldmath$24.5381848$ \\ \hline
$35	$ & $70$ & $	10828	$ & $1.71085307	$ & $1.35828202	$ & $90	$ & $65.4499	$ & $15572.5	$ & \boldmath$28.0148349	$ & \boldmath$25.7532761$ \\ \hline
$36	$ & $66	$ & $8923	$ & $0.63554128	$ & $0.38864304		$ & $91	$ & $	61.9437	$ & $14810.5	$ & \boldmath$25.0950896	$ & \boldmath$22.9313576$ \\ \hline
$37$ & $62$ & $11401$ & $3.91545238$ & $	3.33749576$ & $92$ & $45.587$ & $15786.1$ & \boldmath$45.8401751$ & \boldmath$42.7326945$ \\ \hline
$38$ & $85$ & $6620$ & $1.91263726$ & $2.20926621$ & $93$ & $77.4179$ & $10049.9$ & $0.6859538$ & $0.4843814$ \\ \hline
$39$ & $40$ & $3826$ & $4.6532452$ & $4.60719581$ & $94$ & $77.7532$ & $17395.4$ & \boldmath$33.8241598$ & \boldmath$31.6490305$ \\ \hline
$40$ & $48$ & $8192$ & $2.80768265$ & $2.42994117$ & $95$ & $78.3502$ & $14604$ & \boldmath$15.0196969$ & \boldmath$13.8092691$ \\ \hline
$41$ & $56$ & $6414$ & $1.4363055$ & $1.25608755$ & $96$ & $82.5976$ & $16730.6$ & \boldmath$26.1202574$ & \boldmath$24.5321761$ \\ \hline
$42$ & $58$ & $8900$ & $1.595225$ & $1.23368953$ & $97$ & $29.585$ & $7585.24$ & \boldmath$13.6407061$ & \boldmath$13.0553409$ \\ \hline
$43$ & $69$ & $9406	$ & $0.59520748$ & $0.36094847$ & $98$ & $81.8289$ & $15146.1$ & \boldmath$16.530755$ & \boldmath$15.3924946$ \\ \hline
$44$ & $107$ & $7235$ & $5.04981971$ & $6.06154322$ & $99$ & $95.1694$ & $14773.8$ & \boldmath$10.6986445$ & \boldmath$10.6970408$ \\ \hline
$45$ & $76$ & $8177$ & $0.11949718$ & $0.12548402$ & $100$ & $84.5344$ & $14343.4$ & \boldmath$11.4742294$ & \boldmath$10.7546526$ \\ \hline
$46$ & $97$ & $7684$ & $2.26814245$ & $2.89569778$ & $101$ & $67.6388$ & $15552$ & \boldmath$26.4485017$ & \boldmath$24.307586$ \\ \hline
$47$ & $61$ & $3306$ & $5.17401601$ & $5.15904676$ & $102$ & $77.39$ & $14528.7$ & \boldmath$15.0249637$ & \boldmath$13.7758252$ \\ \hline
$48$ & $87$ & $	8026$ & $0.67345878$ & $0.96969195$ & $103$ & $98.0156$ & $11088.2$ & $1.5684548$ & $2.5439947$ \\ \hline
$49$ & $57$ & $4892$ & $2.55661816$ & $2.46667548$ & $104$ & $71.9236$ & $16326.3$ & \boldmath$29.2832368$ & \boldmath$27.1141681$ \\ \hline
$50$ & $74$ & $8692$ & $0.0861478$ & $0.01293277$ & $105$ & $54.7329$ & $14310.8$ & \boldmath$26.7631647$ & \boldmath$24.4888202$ \\ \hline
$51$ & $67$ & $11272$ & $2.75867222$ & $2.29249393$ & $106$ & $55.5716$ & $15023.1$ & \boldmath$31.1218466$ & \boldmath$28.6249365$ \\ \hline
$52$ & $76$ & $8640$ & $0.04272029$ & $0.01786391$ & $107$ & $108.87$ & $17778.2$ & \boldmath$23.4963946$ & \boldmath$23.9989918$ \\ \hline
$53$ & $55$ & $6989$ & $1.3873851$ & $1.15916632$ & $108	$ & $76.8184$ & $	12283.3$ & $5.6463346$ & $4.9564939$ \\ \hline
$54$ & $65$ & $9352$ & $0.97500274$ & $0.67252189$ & $109$ & $85.8983$ & $17301.1$ & \boldmath$28.5262119$ & \boldmath$26.9984656$ \\ \hline
$55$ & $81$ & $12825$ & $3.17592081$ & $2.89116518$ & $110$ & $98.0195$ & $12851.3$ & $4.114018$&$4.78631$ \\ \hline
\end{tabular}
}
\end{table}
\section{conclusion}\label{sec5}
In manufacturing processes, several procedures release a multivariate data set. They reflect the quality of some different product specifications. In statistical quality control, the main goal is to monitor such data, but their distribution is unknown. So, it is hard to define fitting control limits to the process. There are some traditional methods to deal with these situations, but there is a strong assumption around distribution. This assumption makes them weak in performance, and far from efficient to detect small shifts in several processes. In this paper, we find a joint density function and then get a control chart. The fundamental way is to link the maximum entropy principle and copula functions. The idea of inserting the copula is to preserve the original dependency of the data and transfer it to the estimated distribution. In this regard, we apply $T^2$-Hotelling statistics while dealing with multivariate data. It is common to approximate $T^2$-Hotelling distribution via Fisher distribution when the data set has a Normal distribution, but it is not true in general. By the aim of this paper, we calculate a $UCL$ without $T^2$-Hotelling distribution through the maximum copula entropy.\\
In the end, we add a simulation study and find copula functions based on some dependency measures such as Spearman's rho and Blest. We draw some maximum copula functions and their corresponding density functions to show the dependency effect. The goal is to get the unknown multivariate distribution of manufacturing process data whose variables are dependent. Then, we would like to find a statistical quality control limit according to this distribution for all data. The problem with classical methods is that variable dependencies are not paid attention to. We use average run lengths as well to show the ability of our manner. So, $ARL_0$s and $ARL_1$s are provided, which display the capability in some small changes. Two practical data studies are considered here to show the performance in reality. We explain the details in the first example to show how to calculate the control limit in this method. Also, in this part, the maximum copula entropy performed well when we compared in-control data with out-of-control data. The overall implementation of the introduced control chart is shown by some simulation situations and some empirical data based. The new control chart is offered for some production processes having strict standards.
\clearpage

\bibliographystyle{amsplain}

%
\end{document}